\documentclass{ws-rv9x6}
\usepackage{subfigure}   
\usepackage{ws-rv-thm}   
\usepackage{ws-rv-van}   
\makeindex

\usepackage{bm}
\usepackage{tabularx}
\usepackage{amssymb}
\usepackage{color}
\usepackage{mathtools}
\usepackage{tikz}
\usepackage{url}
\usepackage{hyperref}
\usepackage{cite}
\usetikzlibrary{decorations.pathmorphing,decorations.markings}
\usetikzlibrary{decorations.pathreplacing}
\usetikzlibrary{positioning, calc, shapes}
\usetikzlibrary{shapes.geometric, arrows}
    \tikzstyle{startstop} = [rectangle, rounded corners, minimum width=2cm,text centered, draw=black, fill=red!30]
    \tikzstyle{check} = [rectangle, minimum width=2.0cm, text centered, draw=black, fill=blue!30]
    \tikzstyle{iout} = [trapezium, trapezium left angle=-70, trapezium right angle=-110, minimum width=3cm, text centered, draw=black, fill=blue!30]
    \tikzstyle{io} = [trapezium, trapezium left angle=70, trapezium right angle=110, minimum width=3cm, minimum height=1cm, text centered, draw=black, fill=blue!30]
    \tikzstyle{nnpdf} = [rectangle, minimum width=3cm, minimum height=1cm, text centered, draw=black, fill=orange!30]
    \tikzstyle{n3py} = [rectangle, minimum width=3cm, minimum height=1cm, text centered, draw=black, fill=green!30]
    \tikzstyle{n3cpp} = [rectangle, minimum width=3cm, minimum height=1cm, text centered, draw=black, fill=blue!30]
    \tikzstyle{procblue} = [rectangle, minimum width=3cm, minimum height=1cm, text centered, draw=black, fill=blue!30]
    \tikzstyle{fitted} = [rectangle, minimum width=5cm, minimum height=1cm, text centered, draw=black, fill=red!30]
    \tikzstyle{fixed} = [rectangle, minimum width=5cm, minimum height=1cm, text centered, draw=black, fill=blue!30]
    \tikzstyle{arrow} = [thick,->,>=stealth]

\definecolor{darkgreen}{rgb}{0.0, 0.5, 0.13}

\begin{document}

\chapter{Parton distribution functions}

\author{Stefano Forte and Stefano Carrazza}

\address{Tif Lab, Dipartimento di Fisica, Universit\`a di Milano and\\ INFN, Sezione di Milano,\\ Via Celoria 16, I-20133 Milano, Italy}

\begin{abstract}
We discuss the determination of the parton substructure of hadrons
by casting it as a peculiar form of pattern recognition problem in
which the pattern is a probability distribution, and 
we present the way this problem has been tackled and solved. Specifically, we
review the NNPDF approach to PDF determination, which is based on the
combination of a 
Monte Carlo approach with neural networks as basic underlying
interpolators. We discuss the current NNPDF methodology, based on
genetic minimization, and its validation through closure testing. We
then present recent developments in which a hyperoptimized
deep-learning framework for PDF determination is being
developed, optimized, and tested. 
  
\end{abstract}
\bigskip
\bigskip
\bigskip

\begin{center}
  
  Sumbitted for review.\\To appear in the volume\\ {\it Artificial Intelligence for Particle Physics}, World Scientific Publishing.
\end{center}
\vfill
\begin{flushright}
  TIF-UNIMI-2020-23
\end{flushright}

\clearpage

\body

\tableofcontents

\clearpage

\section{Introduction}
\label{sec:intro}

\begin{figure}
    \center
    \includegraphics[width=1.0\textwidth]{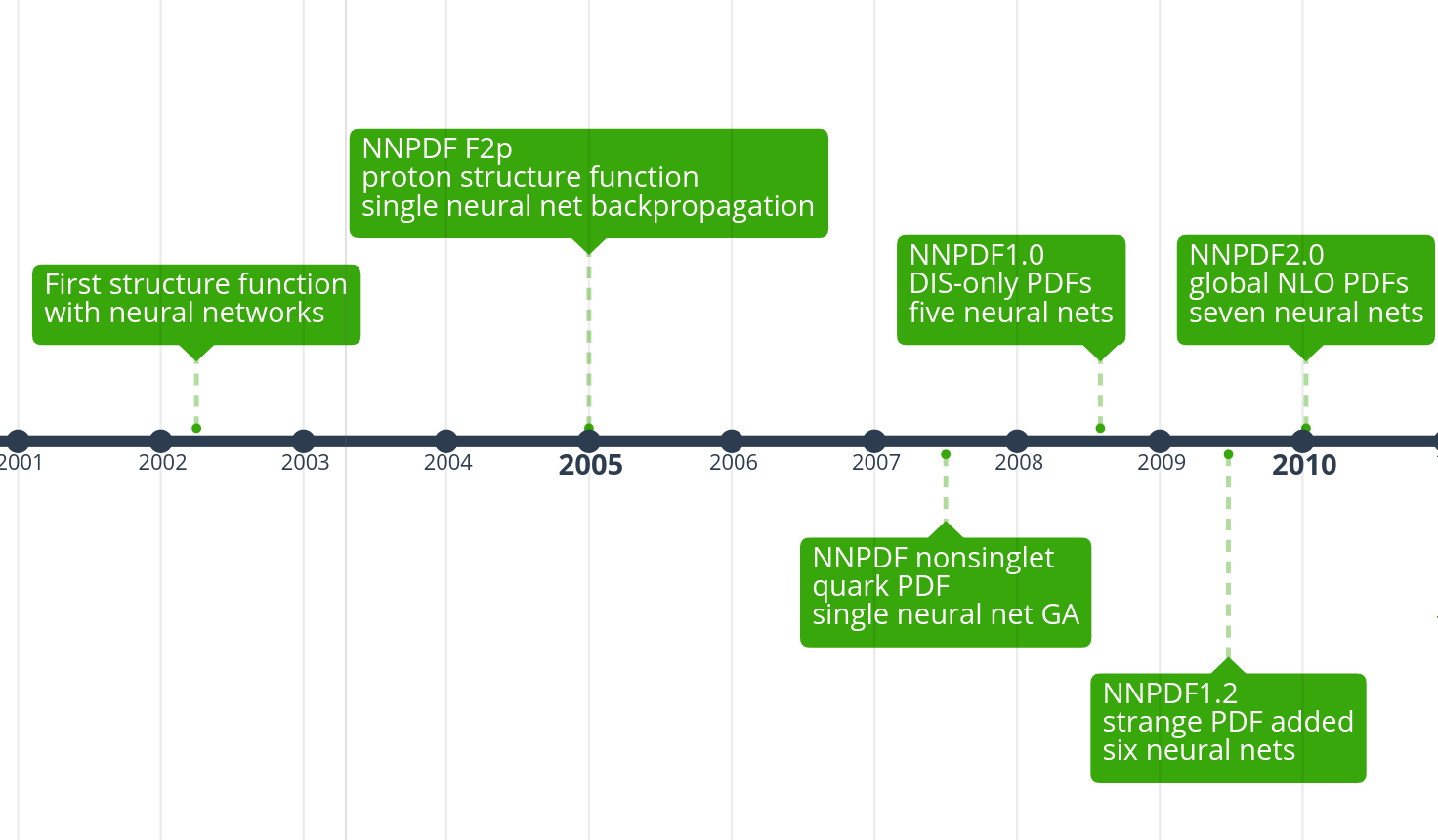}\\
    \includegraphics[width=1.0\textwidth]{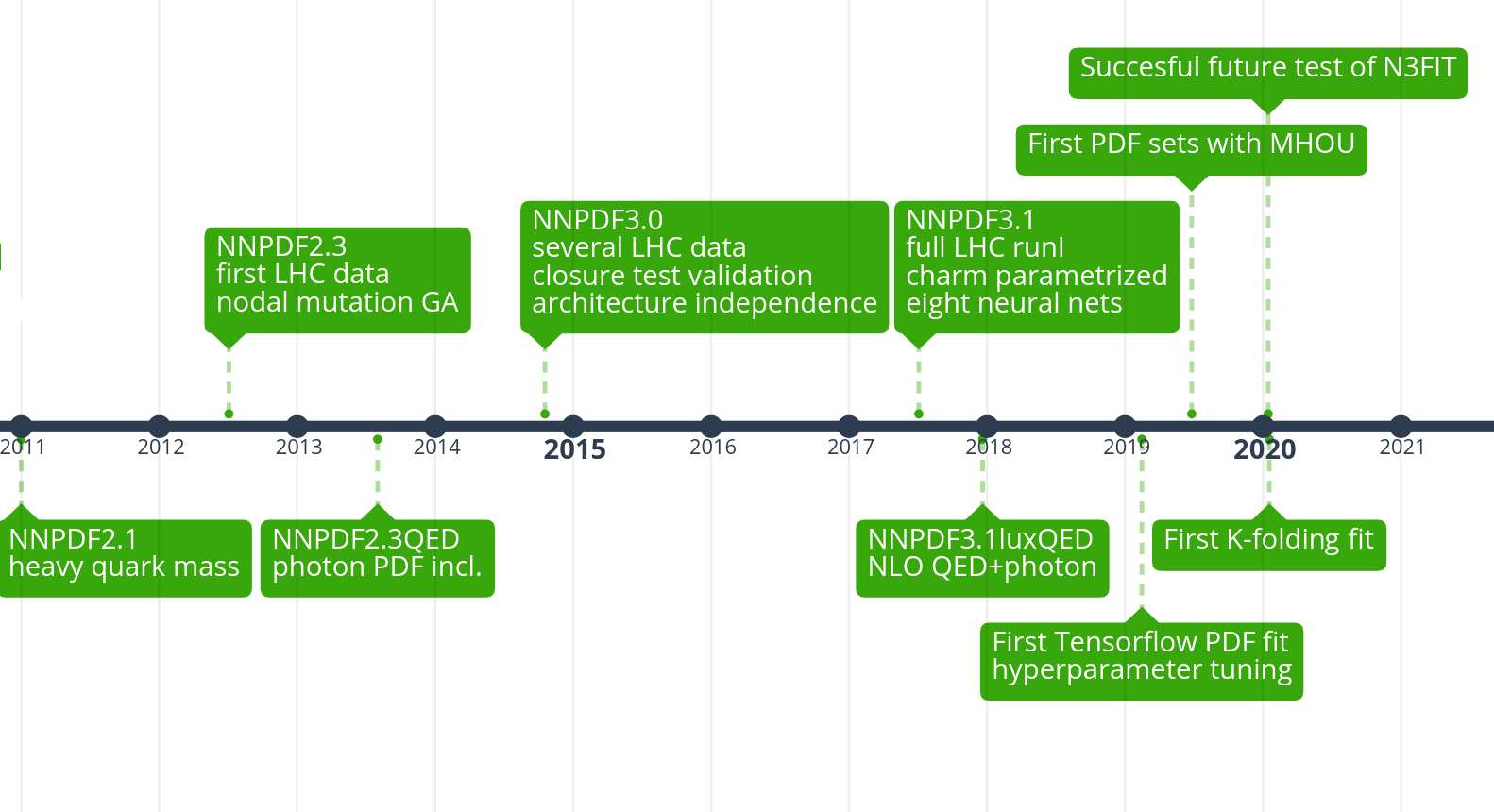}
    \caption{Timeline for the development of PDFs based on AI techniques.}
    \label{fig:timeline}
\end{figure}

The determination of the parton substructure of the nucleon is
essentially a  pattern recognition problem: given an unknown underlying
function that maps input instances to actually realized outcomes, use
a set of data to infer the function itself. However,  the
determination of parton distributions (PDFs, henceforth)
determination differs from standard pattern recognition problems
(such as, say, face detection) in many peculiar and perhaps
unique relevant aspects. Also, whereas the first PDF
determinations have been performed around forty-five years
ago~\cite{McElhaney:1973nj,Kawaguchi:1976wm,DeRujula:1976tz,Johnson:1976xc,Gluck:1976iz,Hinchliffe:1977jy}
it was only
recognized less than twenty
years ago~\cite{Forte:2002fg,f2p,DelDebbio:2007ee,Ball:2008by,Ball:2010de}
that AI techniques could be used for PDF determination~\ref{fig:timeline}).

In this section we will first briefly review what the problem of PDF
determination
consists of, in which sense it can be viewed as a pattern recognition
problem,  and the peculiarities that
characterize it. We will then briefly summarize the NNPDF approach to
PDF determination, which is the only approach in which the problem has been
tackled using AI techniques.

In Section~\ref{sec:statart} we will provide a more detailed
discussion of the NNPDF tool-set used for the determination of current
published PDF
sets i.e. up to NNPDF3.1~\cite{Ball:2017nwa}. We will
specifically discuss the use of neural nets as PDF
interpolants, PDF training using genetic minimization and
cross-validation, and the validation methodology based on closure
testing.
In Section~\ref{sec:PDFdl} we will then turn to a methodology that
is currently being developed for future PDF determinations, which
updates the standard AI tools used by NNPDF to more recent machine
learning methods, relying on deterministic minimization, model
optimization (hyper-optimization) and more powerful and detailed
validation techniques.

\subsection{PDF determination as an AI problem}
\label{sec:pdfsai}

PDFs encode the structure of strongly-interacting
particles or nuclei, as probed in high-energy collisions. A review of
the underlying theory  is beyond the scope of this work, and the
reader is referred to standard textbooks~\cite{Ellis:1991qj}, summer
school lecture notes~\cite{Forte:2010dt} and recent specialized
reviews~\cite{Gao:2017yyd,Ethier:2020way} for more detailed
discussions. Here it will suffice to say that a generic observable,
such as the total cross section $\sigma_X(s,M_X^2)$ for
 a ``hard''
(i.e., perturbatively
computable in QCD)
physical process in a collision
between two hadrons (such as two protons at the
LHC)  has the structure
\begin{equation}\label{eq:fact}
\sigma_X(s,M_X^2) = \sum_{a,b} \int_{x_{\rm min}}^1 dx_1 \, dx_2 \, f_{a/h_1}(x_1,M_X^2)
f_{b/h_2}(x_2,M_X^2){\hat \sigma_{ab\to X}}\left(
x_1x_2 s,M_X^2\right).
\end{equation}
Here $s$ is the (square) center-of-mass energy of the collision (so
$s=(13~{\rm TeV})^2$ at the LHC) and $M_X$ is the mass of the final
state (so $M_X=125~{\rm GeV}$ for Higgs production);
$\sigma_X$ is the measurable cross section, observed in proton-proton
interactions (hadronic cross section, henceforth),
while ${\hat \sigma_{ab\to X}}$ is
the computable
cross section, determined in perturbation theory from
the interaction of two incoming partons, i.e. quarks and gluons $a$
and $b$ (partonic cross section,
henceforth).

In Eq.~(\ref{eq:fact})  $f_{a/h_1}$, $f_{b/h_2}$ are the PDFs: they
provide information on the probability of
extracting a parton of kind $a$, $b$
(up quark, up antiquark, etc.) from incoming hadrons $h_1$,
$h_2$. Note that PDFs
are not quite  probability densities, first because they are not
functions but rather distributions (like the Dirac delta),
and also, they are not positive
definite.
The PDFs
are a universal property of the given hadron: e.g., the proton PDFs are
the same for any process with a proton in the initial state.
They
depend on $x$, which can be viewed as the fraction of the momentum of
the incoming hadron carried by the given parton, so $0\le x\le 1$,
and on the scale
$M^2_X$.
The dependence on $M_X^2$ is computable in perturbation
theory, just like the partonic cross section
${\hat \sigma_{ab\to     X}}$, and it is given as a set of
integro-differential equations, having as initial conditions the set of PDFs
at some reference scale $Q_0$.

The
dependence of the PDFs on $x$ would be computable if one was able to
solve QCD in
the nonperturbative domain: i.e., if it was possible to compute  the
proton wave function from first principles. This is of course not the
case, other than through
lattice simulations~\cite{Lin:2020rut}. Hence, in principle, PDFs for
any given hadron at some reference scale $Q_0$
are a set of well-defined  functions of $x$, namely $f_{a/h}(x,Q_0^2)$, which
depend on the single free parameter of the theory, the strong
coupling (and, for  heavy quark PDFs,  the heavy quark
masses). We know that these functions exist, but we do not know what
they are: at present,  they can only be determined by comparing
cross sections of the form Eq.~(\ref{eq:fact}) for a wide enough set
of observables for which the
hadronic cross section is measured with sufficient precision, and the
partonic cross section is known with sufficient accuracy (i.e. to high
enough perturbative order in QCD, including electroweak corrections,
etc.).

The traditional way the problem has been approached is by postulating
a particular functional form for the $x$ dependence of the
PDFs at a reference scale $Q_0$, given in terms of a set of free parameters;
determining the PDFs at all other scales $Q$ by solving perturbative
evolution equations;  and determining the
free parameters by fitting to the data. The standard choice,
adopted since the very first attempts~\cite{McElhaney:1973nj} is
\begin{equation}\label{eq:simfunct}
  f_i= x^{\alpha_i} (1-x)^{\beta_i},
  \end{equation}
where now $i$ collectively indicates the type of parton and of parent hadron.
This functional form is suggested by theory arguments (or
perhaps prejudice) implying that PDFs should display power-like behavior as
$x\to0$ and as $x\to1$  (see
e.g. Ref.~\cite{Roberts:1990ww}){}. Note that,  even if this were true,
there is no reason to believe that this behavior should hold for all
$x$, and thus, given that only a finite range in $x$ is experimentally
accessible
(currently roughly $10^{-4}\lesssim x \lesssim 0.5$), it is unclear
that this functional form
should apply at all in the observable region.
Furthermore, from the equations which govern the
$Q^2$ dependence of the PDFs, it is easy to see that
even if the PDF takes the form of Eq.~(\ref{eq:simfunct}) at some
scale, this form is not preserved as the scale is varied:
specifically, it is corrected by $\ln x$ terms as $x \to0$, and by $\ln(1-x)$
terms as $x\to1$.

The fact that the simple functional form Eq.~(\ref{eq:simfunct}) is
too restrictive has been rapidly recognized, and
more and more elaborate  functional forms
have been adopted in more recent PDF determinations. For example,
the gluon PDF of
the proton was parametrized in the CTEQ5~\cite{Lai:1999wy} PDF set as
\begin{equation}\label{eq:ct5g}
  xg(x,Q_0^2)=A_0x^{A_1} (1-x)^{A_2}(1+A_3 x^{A_4})
\end{equation}
 and in the CT18 PDF set~\cite{Hou:2019efy} as
\begin{align}\label{eq:ct18g}
g(x,Q=Q_0) &= x^{a_1-1}(1-x)^{a2}\left[a_3         (1-y)^3 +
a_4   3 y   (1-y)^2 +
a_5   3 y^2 (1-y)   +
y^3\right];\nonumber\\
 &y=\sqrt x;\quad a_5 = ( 3 + 2 a_1 )/3.
\end{align}
Issues related to postulating a fixed functional form for PDFs were made apparent when a
determination of the uncertainties on the PDFs was first
attempted~\cite{Stump:2001gu,Pumplin:2001ct,Martin:2002aw}. Namely,
uncertainties on the fit parameters
determined by least-squares and standard error
propagation  turned out to be smaller by about
one order of magnitude than one might reasonably expect by looking at
the fluctuation of best-fit values as the underlying dataset was
varied. This led to the peculiar concept of ``tolerance'', namely, an
a-posteriori
rescaling factor of uncertainties. It is debatable how much of
the need for such a rescaling is related to the bias introduced by the
choice of a particular functional form. However, a not uncommon
occurrence is that  addition of new data,
leading to a more extended parametrization (such as
Eq.~(\ref{eq:ct18g}) in comparison to Eq.~(\ref{eq:ct5g})) would lead to
an {\it increase} in uncertainties. This suggests that the
more restrictive parametrization might well be biased.

In 2002 it was first suggested~\cite{Forte:2002fg} that these
difficulties may be overcome by addressing the problem of PDF
determination by means of  a standard AI tool, neural
networks. The basic underlying intuition is that neural networks
provide a universal interpolating function, and that by choosing a
sufficiently redundant architecture any functional form can be
accommodated in a bias-free way, while avoiding overtraining through
suitable training methods, as we will discuss in
Sections~\ref{subsec:cv},~\ref{sec:kfold} below.
This first suggestion was gradually developed into a
systematic methodology for PDF determination through a series of
intermediate steps (see
Figure~\ref{fig:timeline}) involving, on the methodological side,  a
number of subsequent improvements, to be discussed below, and a set of
validation and testing techniques. The more recent
successors NNPDF3.0~\cite{Ball:2014uwa} and
NNPDF3.1~\cite{Ball:2017nwa} of the first PDF set developed using this
methodology
(NNPDF1.0~\cite{Ball:2008by})
are
currently the most widely cited PDF sets.

It should now be clear in which sense PDF determination can be viewed
as a  pattern recognition problem, and what are its peculiar
features. As in standard pattern recognition, the main goal is to
determine a set of unknown underlying functions from data instances,
with almost no knowledge of their functional form (other than loose
constraints of integrability with an appropriate measure, smoothness,
etc.).
Unlike in the simplest pattern
recognition problems, the functions provide continuous output (i.e. the features
to be recognized are continuous), and
data are not directly instances of the functions to
be determined. Hence, one cannot associate an input-output pair
to an individual data point. Rather, as apparent from Eq.~(\ref{eq:fact}),
each datapoint provides an output which
depends in a nonlinear way on the full set of functions evaluated at
all input values, which are integrated over from some minimum
$x_{\rm min}$ (depending on the particular observable and the values of $s$
and $M_X^2$).
This is of course common to  more complex pattern
recognition problems, such as in computer vision.

There are however two peculiarities in PDF determination which set it
apart from most or perhaps all other applications of AI. The first is
that the quantities which one is trying to determine, the PDFs, are
probability distributions of observables, rather than being
observables themselves. This follows from the fact that, due to
the quantum nature of
fundamental interactions, individual events (i.e. measurement
outcomes) are stochastic, not deterministic.
Even if the PDF were known exactly to
absolute accuracy, the cross section would just express the probability
of the observation of an event, to be determined through repeated
measurements. The PDFs are accordingly probability
distributions. The goal of PDF determination is to determine the
probability distribution of PDFs: hence, in PDF determination one
determines a probability distributions of probability distributions,
i.e. a probability functional.

The
second peculiarity is that in order for a PDF determination to be useful as an
input to physics predictions, full knowledge of PDF correlations is
needed. In fact,  PDF uncertainties are typically a
dominant source of uncertainty in predictions for
current and future high-energy experiments~\cite{Azzi:2019yne}. But
the uncertainty on each
particular PDF at a given $x$ value, $f_i(x,Q_0^2)$ is correlated to the
uncertainty on any other PDF at a different $x$ value
$f_j(x',Q_0^2)$, and this correlation must be accounted for in order
to reliably estimate PDF uncertainties~\cite{Bagnaschi:2019mzi}.
Hence, PDF determination also requires the determination
a covariance matrix of
uncertainties in the space of
probability distributions: namely,
a covariance matrix functional.

The NNPDF approach to PDF determination
tackles this problem using AI tools, as we discuss in the next section.

\subsection{The NNPDF approach}
\label{sec:NNPDFappr}

\begin{figure}
    \center
    \includegraphics[width=1.0\textwidth]{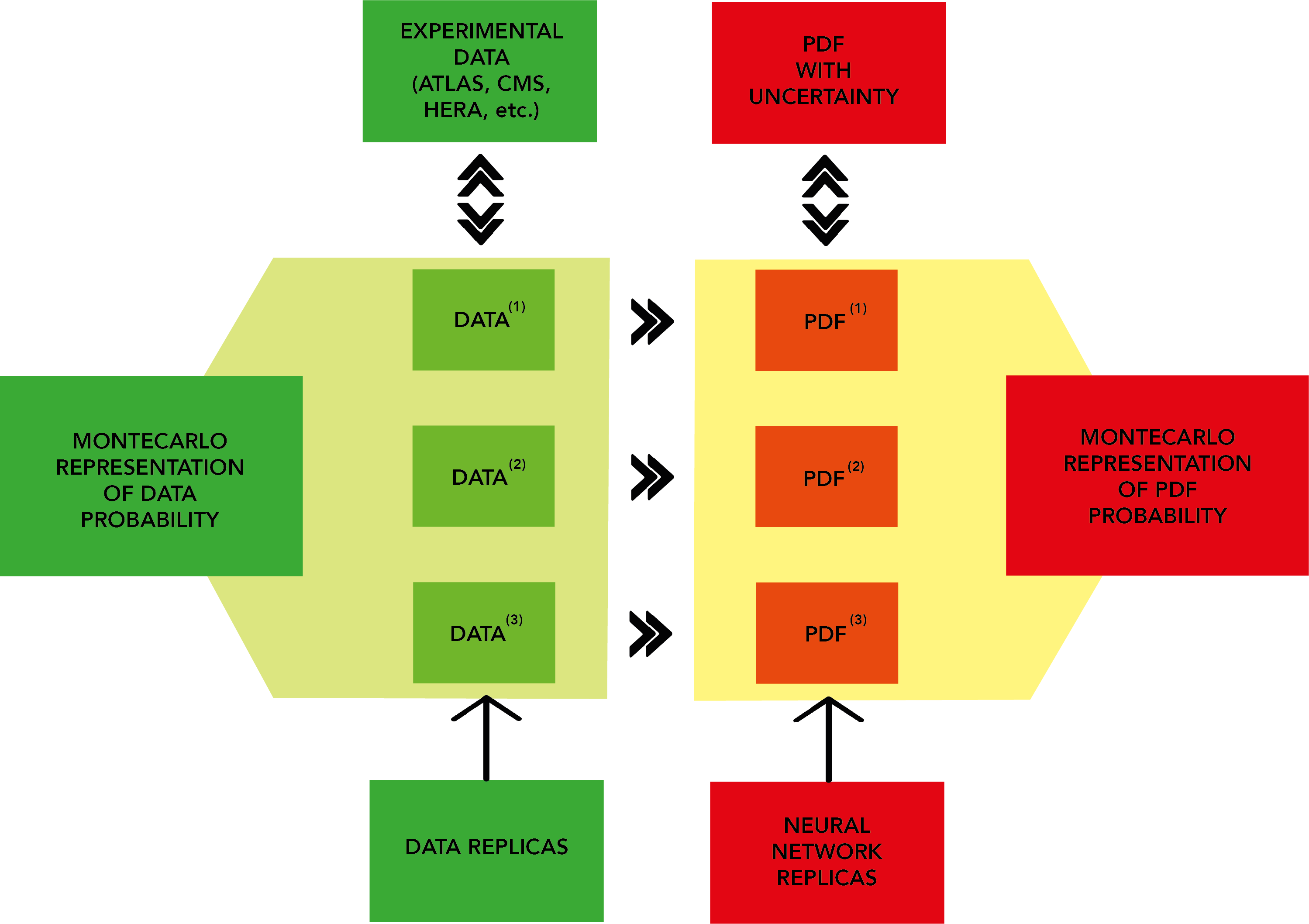}
    \caption{Schematic representation of the NNPDF methodology.}
    \label{fig:methodology}
\end{figure}

As seen in Sect.~\ref{sec:pdfsai} the NNPDF methodology has the
goal of determining the probability distribution of a set of
functions, which in turn are related to
the probability distributions of quantum events (the
emission of a parton from a parent hadron) which provide the input to
the computation of predictions for (discrete) experimental
measurements. The methodology is based on two distinct ingredients:
the use of a Monte Carlo representation for the probability
distributions, and the use of neural networks as unbiased underlying
interpolating functions. It is schematically represented in
Figure~\ref{fig:methodology}.

The Monte Carlo representation provides a way of breaking down the 
problem of determining a probability in a space of functions into an
(in principle infinite) set of problems in which a unique best-fit set
of functions is determined. The basic idea is to turn the input
probability distribution of data into a Monte Carlo
representation. This means that  the input data and correlated
uncertainties are viewed as a probability distribution (typically, but
not necessarily, a multigaussian) in the space of data, such that the
central experimental values correspond to the mean and the correlated
uncertainties correspond to the covariance of any two data. The Monte
Carlo representation is obtained by extracting a set of replica instances from
this probability distribution, in such a way that, in the limit of
infinite number of replicas, the mean and and covariance over the
replica sample reproduce the mean and covariance of the underlying
distributions. In practice the number of replicas can be determined a
posteriori by verifying that mean and covariance are reproduced to a
given target accuracy.

A best-fit PDF (or rather, PDF set: i.e. one function $f_i(x,Q_0^2)$
for each distinct  type of parton $i$) is then determined for each
data replica, by minimization of a suitable figure of merit. Neural
networks are used to represent the PDFs, with the  value of $x$ as
input, and the value of the PDF as an output (one for each PDF).
Note that the fact that the data only depend
indirectly on the input functions to be determined (the PDFs) is
immaterial from the point of view of the general methodology.
Indeed, the problem has been reduced to that of determining the
optimal PDFs for each input data replica, namely, to standard training
of neural networks. However, the fact that the PDF is not trained to
the data directly will have significant implications on the nature of
PDF uncertainties, on their validation, and on the optimization of PDF training,
as we will discuss more extensively in Sections~\ref{sec:valid},~\ref{sec:newdet},~\ref{sec:hyperopt}.

The output of the process is a set of PDF replicas, one for each
data replica. These provide the desired representation of the
probability density in the space of PDFs. Specifically, central
values, uncertainties and correlations can be computed doing
statistics over the space of PDF replicas: the best-fit PDF is the
mean over the set of replicas, the uncertainty on any PDF for given
$x$ can be found from  the variance over the replica sample, and the
correlation from the covariance.

The remaining methodological problems are how to determine the optimal neural network
parametrization, how to determine the optimal PDF for each replica
(i.e. the optimal neural network training) and how to validate the
results. The way these issues are addressed in the current NNPDF
methodology will be discussed in Section~\ref{sec:statart}, while
current work towards improving and hyperoptimizing the methodology are
discussed in Section~\ref{sec:PDFdl}.

\section{The state of the art}
\label{sec:statart}

The NNPDF methodology, presented  in Sect.~\ref{sec:NNPDFappr},
combines a Monte Carlo approach representation of probability
distributions with neural networks as basic interpolants. Here we
discuss first, the architecture of the neural networks, then their
training, which is achieved by combining genetic minimization with
stopping based on cross validation, and finally the validation of
results through closure testing.

\subsection{Neural networks for PDFs}
\label{sec:nn}

\begin{figure}
    \center
    \includegraphics[width=0.65\textwidth]{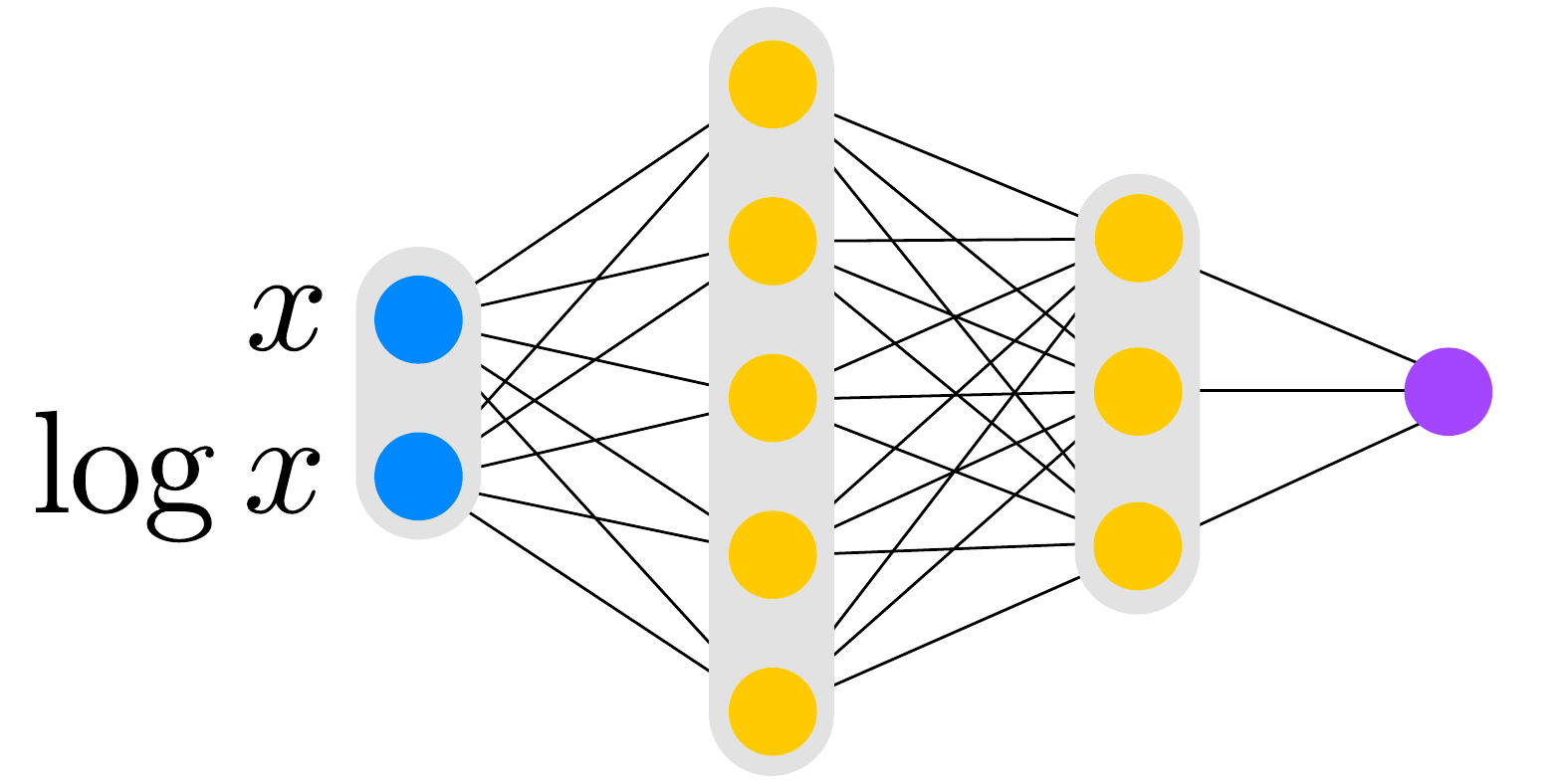}
    \caption{Architecture of the neural networks used for PDF
      parametrization in all available NNPDF sets. Each PDF is
      parametrized by a preprocessed neural network, according to
      Eq.~(\ref{eq:PDFdefinition}). The values of $x$ and $\ln x$ are
      taken as input, and the value of the PDF is given as output. The number of independently
      parametrized  PDFs has increased over time but the architecture
      has remained the same.
    }
    \label{fig:arch}
\end{figure}
In all NNPDF determinations, starting with the proof-of-concept
determination of a single PDF (isotriplet combination) in
Ref.~\cite{DelDebbio:2007ee}{}, up to and including the most recent
global PDF set, NNPDF3.1~\cite{Ball:2017nwa} the PDF architecture has
been unchanged. Namely, PDFs
are parameterized at a reference scale
$Q_0$ and expressed in terms of a set of independent neural networks
 multiplied by a preprocessing factor. Each of these neural
networks consists of a fixed-size feed-forward multi-layer perceptron with
architecture 2-5-3-1 (see Fig.~\ref{fig:arch}).
The only change in subsequent
releases is in the number of independently parametrized PDFs (or PDF
combinations), and thus of independent neural networks: one in the
proof-of-concept Ref.~\cite{DelDebbio:2007ee}{},
five in NNPDF1.0~\cite{Ball:2008by} (up
and down quarks and antiquarks and the gluon), seven from
NNPDF1.1~\cite{Rojo:2008ke}, eight in NNPDF3.1~\cite{Ball:2017nwa}
(up, down, strange quarks and antiquarks; total charm; gluon).

The PDF momentum fraction $x$ enters the input layer nodes as $(x,
\log(x))$, in order to account for the fact that the physical
behavior of PDFs typically has two different regimes in the
physically accessible $10^{-4}\lesssim x \lesssim 0.5$ region: a linear regime
in the region $0.03\lesssim x\lesssim 0.5$ and a logarithmic regime in
the region $10^{-4}\lesssim x \lesssim 0.03$.
The next two hidden layers, with 5 and 3 nodes respectively, use the sigmoid
activation function while the output node is linear. This particular choice of
architecture was originally selected through systematic
manual scans, as being sufficiently redundant
to accommodate the PDF shape in an unbiased way .

The fact  that it was never necessary to update this
initial choice has validated the robustness of this analysis.
Furthermore, in Ref.~\cite{Ball:2008by} it was explicitly checked
that results would be unchanged if the number or nodes in the first
hidden layer was reduced from 5 to 4. In Ref.~\cite{Ball:2014uwa}{},
within a closure test (see Section~\ref{sec:valid} below), it was
checked that results were unchanged if the number of the nodes in the
intermediate layers was increased respectively from 5 to 20 and from 3
to 15, which corresponds to an increase of the number of free
parameters of the neural net by more than one order of magnitude.

The parametrization for each PDF (or independent combination of PDFs) is
\begin{equation}
    x f_i(x,Q_0) = A_i x^{-\alpha_i+1} (1-x)^{\beta_i} {\rm NN}_i(x), \label{eq:PDFdefinition}
\end{equation}
where ${\rm NN}_{i}$ is the neural network corresponding to a given
combination $i$. The quantities which are independently parametrized are
the linear combination of light quark and gluon
PDFs which correspond to eigenvectors of the
PDF $Q^2$ evolution equations, and charm:
$\{g,\,\Sigma,\,V,\,V_3,\,V_8,\,T_3,\,T_8,\,c^+\}$ (see
Refs.~\cite{Forte:2010dt,Ball:2017nwa} for the precise definition). $A_i$ is an overall
normalization constant which enforces sum rules (such as the fact that
the total momentum fractions
carried by all partons must add up to one) and
$x^{-\alpha_i}(1-x)^{\beta_i}$
is a preprocessing factor which controls the PDF behavior at
small and large $x$.

The preprocessing exponents $\alpha_i$ and $\beta_i$ were initially
(NNPDF1.0~\cite{Ball:2008by}) chosen to be fixed, while checking that
no strong dependence of results was observed upon their
variation. As the accuracy of the PDF determination improved, starting
with NNPDF1.2~\cite{Ball:2009mk}, in order
to ensure unbiased results, the exponents were varied. Namely, the
values of $\alpha_i$, $\beta_i$ were randomly selected for each PDF in
each replica,
with uniform distribution within a range fixed for each PDF,
and kept fixed during the minimization of the replica. Effectively,
with reference to Fig.~\ref{fig:methodology}, this means that for each
PDF replicas the PDF parametrization is different, because the
preprocessing function of each PDF is different. The range for each
type of PDF (gluon, up quark, etc) was
initially determined by requiring stability of the fit results, which, starting
with NNPDF2.0~\cite{Ball:2010de} was quantitatively determined by
computing the correlation coefficient between the figure of merit $\chi^2$ (see
Eq.~\eqref{eq:chi2} below) and verifying that it remained small.
Starting with NNPDF3.0~\cite{Ball:2014uwa}, the range is now determined
self-consistently: the effective exponents are computed  for each
independent combination
of PDFs and for each PDF replica, the 68\% confidence level range is
determined for each combination,   the
 fit is repeated with the exponents varied in a range taken equal to twice this
range, and the procedure is iterated until the range stops changing.

As already mentioned, unlike in
many standard regression problems, in which  during the optimization
procedure the
model is compared directly to the training input data, in PDF fits the
data are compared to
theoretical predictions for physical observables  of the form of
Eq.~\eqref{eq:fact}, in which the PDFs $f_i(x,Q^2)$ are in turn obtained by
solving a set of integro-differential equations from the PDFs
$f_i(x,Q_0)$, parametrized at the initial scale.  Hence, the
observable depends on the PDF through  a number of
convolution integrals, between the PDFs at
scale $Q_0$, the evolution factors that take them to scale $Q$ and the
partonic cross sections of Eq.~\eqref{eq:fact}. In practice, the
convolutions are turned into multiplication of pre-computed tables
(FastKernel or FK-tables) by
projecting on suitable basis functions, as discussed
in
Refs.~\cite{Ball:2010de,Bertone:2016lga}{}, see also
Section~\ref{sec:newdet} below.

\subsection{The minimization procedure}
\label{sec:ga}

The optimization procedure implemented in NNPDF consists in minimizing the loss
function
\begin{equation}
    \chi^2 = \sum_{i,j}^{N_{\rm dat}} (D-P)_i \sigma_{ij}^{-1} (D-P)_j,
    \label{eq:chi2}
\end{equation}
where $D_i$ is the $i$-th data point, $P_i$ is
the convolution product between the FastKernel tables for point $i$ and the PDF
model, and $\sigma_{ij}$ is the covariance matrix between data points $i$ and
$j$. The covariance matrix includes both uncorrelated and correlated
experimental statistical and systematic uncertainties, as given by the
experimental collaborations. Multiplicative uncertainties (such as
normalization uncertainties), for which the uncertainty is
proportional to the observable, must be handled through a dedicated
method in order to avoid fitting bias: the $t_0$ method has been
developed~\cite{Ball:2009qv} to this purpose, and adopted from
NNPDF2.0~\cite{Ball:2010de} onward. Theory uncertainties (such as
missing higher order uncertainties) could also be included as
discussed in Refs.~\cite{AbdulKhalek:2019bux,AbdulKhalek:2019ihb} but
this has only been done in preliminary PDF sets so far. Once again, we
stress that
input data are not provided  for the
neural networks, but rather for a complicated functional of the neural
network output.

\subsubsection{Genetic minimization}

The minimization implemented in NNPDF3.1 and earlier releases is based on
genetic algorithms (GA). Given that each PDF replica is completely independent
from each other, the minimization procedure can be trivially
parallelized. Genetic minimization was chosen for a number of reason.
On the one hand, it was felt that
that a deterministic minimization might run the risk of ending up in a
local minimum related to the specific network
architecture. Also, no efficient way of determining the
derivative of the observables with respect to the parameters of the
neural network was available then. In fact, modern, efficient
deterministic minimization methods\cite{ADADELTA,Adam} were not yet
available at the time.
As we will discuss in Section~\ref{sec:newdet} below,
these motivations are no longer valid and deterministic minimization is
now more desirable.

The GA algorithm consists of three main steps: mutation, evaluation and
selection. These steps are performed subsequently through a fixed number of
iterations. The procedure starts with the initialization of the neural network
weights for each PDF flavor using a random Gaussian distribution.
From this initial network, a number of
copies is produced, for which the weights are then mutated
with a suitable rule. The mutations with lowest values of the figure
of merit are selected and the procedure is iterated.

The GA initially adopted was based on point change mutations, in which
individual weights or thresholds in the networks  were mutated at
random, according to a rule of the form
\begin{equation}\label{eq:gaw}
w_i \rightarrow w_i + \eta_i r_i \, ,
\end{equation}
where $w_i$ is the $i$-nth neural network weight or threshold,
$\eta_i$ is a mutation rate
size, $r_{i}$ is a uniform random number within [$-1$, $1$]. A fixed
number of randomly chosen parameters are then mutated for each PDF,
thereby producing a given number of mutants for each generation.
The GA is fully specified by assigning: (i) the number of
mutations for each PDFs; (ii) the mutation rates for each mutation and
for each PDF; (iii) the number of mutants for each generation; (iv) the
maximum number of generations.
The mutation rates were dynamically adjusted as a function of the
number of iterations according to
\begin{equation}\label{eq:eta}
\eta_i=\frac{\eta_i^{(0)}}{N_{\rm ite}^p}.
\end{equation}

Several subsequent versions of this GA have been adopted. In a first
version (NNPDF1.0~\cite{Ball:2008by}), a fixed value of the number of
mutations (two per PDF), of the number of mutants ($N_{\rm mut}=120$)
and of the exponent $p$ ($p=1/3$) of
Eq.~(\ref{eq:eta}) were adopted, with a small maximum number of
generations ($N_{\rm max}=5000$). At a later stage
(NNPDF2.0~\cite{Ball:2010de}) the minimization was divided in two
epochs, with a  transition at $N_{\rm ite}=2500$ generations, and a
larger number  ($N_{\rm mut}=80$) of mutants in the first epoch,
substantially decreased  ($N_{\rm mut}=10$) in the second epoch; also
the exponent $p$ was now randomly varied between 0 and 1 at each generation
 and the maximum number of generations
was greatly increased ($N_{\rm max}=30000$). At a yet later stage
(NNPDF2.3~\cite{Ball:2012cx}) the number of mutations was increased to
three for several PDFs.

Subsequent versions of the GA also involved
various reweighting procedures, in which the contribution of different
datasets to the figure of merit Eq.~\eqref{eq:chi2} was assigned a
varying weight during the training, in order to speed up the training
in the early stages.
In a first implementation~\cite{Ball:2008by}, these
weights were
computed as a ratio of the $\chi^2$ per datapoint for the given
dataset, compared to the $\chi^2$ per datapoint of the worst-fitted
dataset, so that best-fitted dataset would get less weight.
Weights were then switched off when the value of
the figure of merit fell below a given threshold.
In a
subsequent implementation~\cite{Ball:2010de}, the weights were
computed as ratios of the $\chi^2$ to a target $\chi^2$ value for the
given dataset (determined from a previous fit) and only assigned to
datasets for which the fit quality was worse than the target. Weights
were only applied in a first training epoch.

Starting with NNPDF3.0~\cite{Ball:2014uwa}, a GA based on nodal
mutation has been adopted. In nodal mutation,
each node in each network is assigned an independent probability
of being mutated.  If a node is selected, its threshold and all of the weights
are mutated according to Eqs.~(\ref{eq:gaw}-\ref{eq:eta}), with now
$\eta$ fixed, and
$p$ a
random number between $0$ and $1$ shared by all of the weights. The
values $\eta=15$ and mutation probability 15\% per node have been selected
as optimal based on closure tests (see Section~\ref{sec:valid}
below). This algorithm proved to be significantly more efficient (see Figure~\ref{fig:level0}
below) that the previous point mutation: in particular, reweighting is no
longer necessary and it is no longer necessary to have different
training epochs.

\subsubsection{Stopping criterion}
\label{subsec:cv}

The GA presented in the previous Section~\ref{sec:ga} can lead to
overfitting, in which not only the underlying law is fitted, but also
statistical noise which is superposed to it. In order to avoid this, a
stopping
criterion is required. This was implemented since NNPDF1.0 through
cross-validation.
Namely, the data are separated in
a training set, which is fitted, and a validation set, which is not
fitted.
The GA minimizes the $\chi^2$ of the training set, while the $\chi^2$
of the validation set is monitored along the minimization, and the
optimal fit is achieved when the validation $\chi^2$ stops
improving. This means that the fit optimizes the validation $\chi^2$,
which is not fitted. Because statistical noise
is uncorrelated between the training and
validation sets, this guarantees that overfitting of the statistical
noise
is avoided. Note
that more subtle form of overfitting are possible, due to remaining
correlations between training and validation sets: this, and the way
to avoid it, will be discussed in Section~\ref{sec:newval} below.

In PDF fits before NNPDF3.0~\cite{Ball:2014uwa}  this stopping criterion was implemented by
monitoring a moving average of the training and validation $\chi^2$,
and stopping when the validation moving average increased while the
training moving average decreased by an amount which exceeded suitably
chosen threshold values. This was necessary in order to avoid stopping
on a local fluctuation, and it required the tuning of the moving
average and of the threshold values, which was done by studying the
typical fluctuations of the figure of merit. This clearly
introduced a certain arbitrariness.

Since NNPDF3.0~\cite{Ball:2014uwa},  the previous stopping criterion
has been replaced by the
so-called \emph{look-back} method. In this method,  the PDF parametrization
is stored for the iteration where the fit reaches the absolute minimum of the
validation $\chi^2$ within a given maximum number of generations. This
guarantees that the absolute minimum of the validation $\chi^2$ within
the given maximum number of iterations is achieved. The
method reduces the level of arbitrariness introduced in the previous
strategy, but  it requires reaching the maximum number of
iterations for all
replicas, out of which the absolute minimum is determined. This
maximum must be chosen to be large enough that the absolute minimum is
always reached, and it therefore
leads on average to longer training. Adoption of this new stopping has
 been made possible
thanks to greater computing efficiency.

\subsection{Closure tests}
\label{sec:valid}
As mentioned in Section~\ref{sec:intro} a critical issue in PDF
determination
is making sure that PDF uncertainties are
faithful. Therefore, the validation of a PDF set chiefly consists of
verifying that PDF uncertainties accurately reproduce the knowledge of
the underlying true PDFs which has been learnt and stored, together with its
uncertainty,  in the Monte Carlo
replica set through the training procedure. Because the true PDFs are
not known, this can only be done through closure
testing~\cite{demortier}. Namely,  a particular underlying
truth is assumed (in our case: a specific form for the true underlying
PDFs); data are then generated based on this underlying truth; the
methodology is applied to this data; results are finally compared to
the underlying truth.

This exercise was performed for the NNPDF3.0 PDF
set~\cite{Ball:2014uwa}; since the subsequent NNPDF3.1 PDF
set~\cite{Ball:2017nwa} is based on the same methodology, this provides
a validation of the current NNPDF PDF sets. In this Section
we will briefly review
the closure testing methodology and results of
Ref.~\cite{Ball:2014uwa}{}, while the ongoing
validation of the new  methodology of Section~\ref{sec:PDFdl} will be discussed
in Section~\ref{sec:qual} below.

In this closure test,  data were generated by
assuming that the underlying PDF has the form of the MSTW08 PDF
set~\cite{Martin:2009iq}, and then generating a dataset identical to
that used for the NNPDF3.0 PDF determination (about 4000 data points)
but computing the hadronic cross sections
using Eq.~\ref{eq:fact} with these PDFs
adopted as input and the partonic cross sections  determined using NLO
QCD theory. Clearly, the exact form of the theory is immaterial if the
same theory is used to generate the data and then to fit them, in such
a way that only the fitting methodology is being tested. The
independence of result on the particular choice of underlying truth
can be explicitly tested by repeating the procedure with a different
choice for the underlying PDF.

Besides providing a validation of
the NNPDF methodology, the closure test also allows for an
investigation of the sources of PDF uncertainty in a controlled
setting. To this purpose, three sets of closure testing data were
generated in Ref.~\cite{Ball:2014uwa}{}.
The first set (``level 0'')
consists of  data generated with no uncertainties. This would
correspond to a hypothetical case in which there are no experimental
statistical or systematic uncertainties, so all data correspond to the
``truth'', with vanishing uncertainty.
A second set of data (``level 1'') is generated by assuming the
probability distribution which corresponds to the published
experimental covariance matrix. These data correspond to a
hypothetical set of experimental results for which the experimental
covariance matrix is exactly correct. A final set of data (``level
2'') is generated by taking the level 1 data as if they were actual
experimental data, and then applying to them the standard NNPDF methodology,
which, as discussed in Section~\ref{sec:NNPDFappr} (See
Figure~\ref{fig:methodology}) is based on producing a set of Monte
Carlo replicas of the experimental data: the level 2 data are then the
Monte Carlo replicas produced out of the level 1 data, as if the latter
were actual experimental data.

\begin{figure}
    \center
    \includegraphics[width=1.0\textwidth]{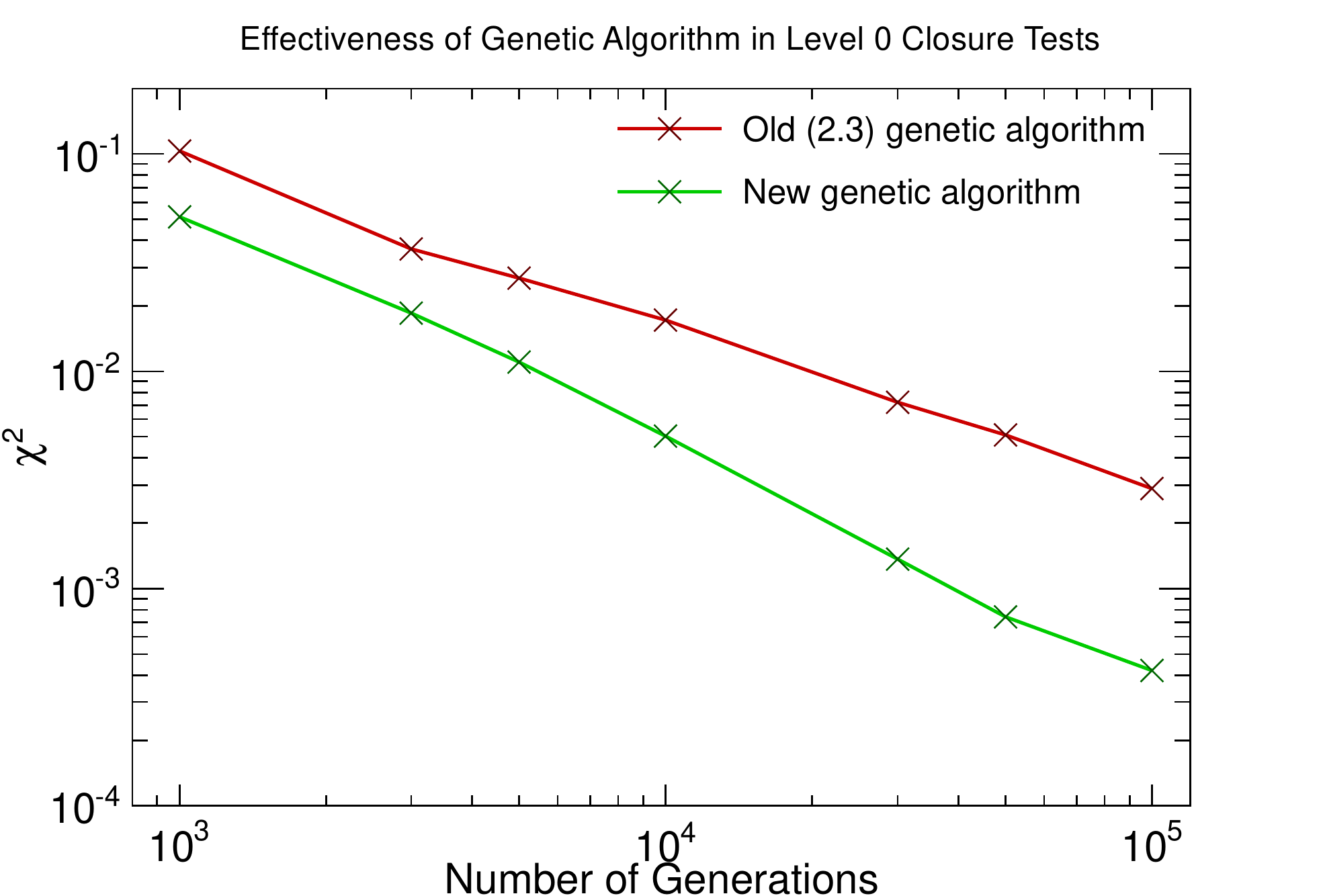}
    \caption{The normalized figure of merit computed for the average
      over PDF replicas vs. the number of
    generations of the genetic algorithm for two different GA
    implementations, in a test case in which the figure of merit
    vanishes asymptotically.}
    \label{fig:level0}
\end{figure}
A first very simple test consists of fitting level~0 data, and
computing the figure of merit
($\chi^2$ per datapoint) as the training proceeds. Because these data
have no uncertainty, a perfect fit with $\chi^2$ is in principle
possible.
Results are shown
in Figure~\ref{fig:level0} for the two implementations of the
minimization algorithm adopted in Refs.~\cite{Ball:2012cx} (NNPDF2.3)
and \cite{Ball:2014uwa} (NNPDF3.0) and discussed in
Section~\ref{sec:ga}. Two sets of conclusions may be drawn from his
plot. First, it is clear that the methodology is general and powerful
enough to reproduce the underlying data: the figure of merit can be
made arbitrarily small, which means that
with vanishing experimental uncertainties, the data can be fitted with
arbitrarily
high accuracy. Second, it is possible to determine the dependence of
the figure of merit on the training length, and specifically compare
different minimization algorithms. Interestingly,
Figure~\ref{fig:level0} shows that for the two GAs of
Section~\ref{sec:ga} the figure of merit follows a power law:
$\chi^2\sim \frac{1}{N^\lambda}$. Furthermore, it is clear that  the
value of $\lambda$ is rather larger
(faster convergence) for the NNPDF3.0 GA, based on nodal mutation
(recall Section~\ref{sec:ga}),
in comparison to the previous NNPDF2.3 GA implementation.

\begin{figure}
    \center
    \includegraphics[width=.48\textwidth]{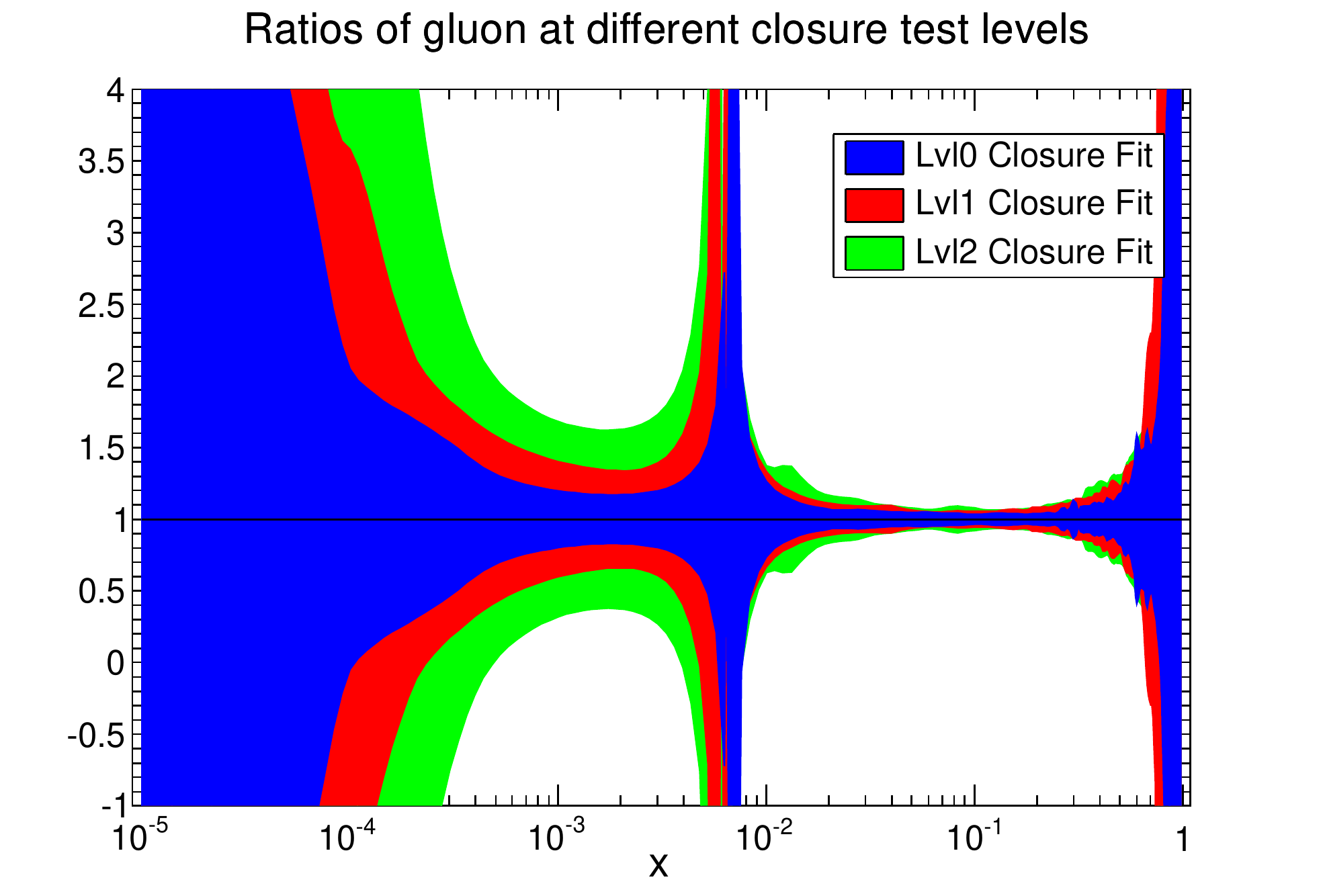}
    \includegraphics[width=.48\textwidth]{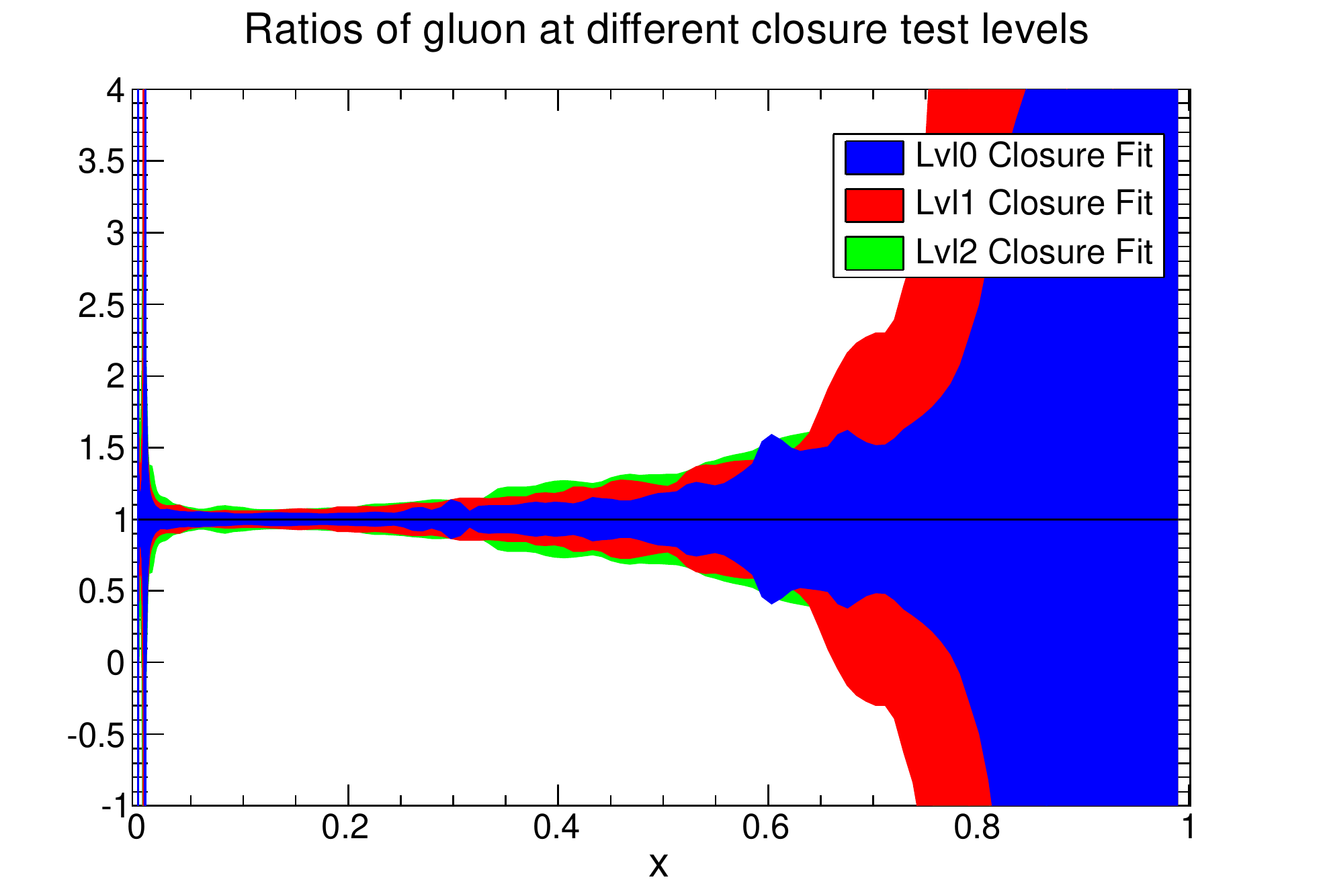}
    \caption{The 68\% confidence level uncertainty bands for the gluon
      PDF determined using level~0, level~1 and level~2 closure test
      data (see text). Results are shown vs. $x$ at the PDF
      parametrization scale  on a
      logarithmic (left) and linear (right) scale.}
    \label{fig:level012}
\end{figure}
A second test compares the uncertainty on PDFs which is
found when fitting respectively to level 0, level 1 and level 2
data. Results are shown for the gluon in Fig.~\ref{fig:level012}:
68\% confidence levels are shown for fits
to level~0, level~1 and level~2 data.
The plot has various
implications. The first observation is that, as
discussed in Section~\ref{sec:intro} the data constrain the PDFs only
in a limited $10^{-2}\lesssim x\lesssim 0.5$ range (``data region'', henceforth). Outside that range
the uncertainty grows very large, and in the absence of experimental
information it is essentially arbitrary.

Coming now to the region where the experimental information is
concentrated, note that when fitting level~0 and level~1 data the
same datapoints are fitted over and over again, yet a spread of
results is found. In the case of  level~0
data we know from Figure~\ref{fig:level0} that
the figure of merit on datapoints essentially vanishes (i.e., the fit goes
through all datapoints with zero uncertainty).   This then means that
this unique minimum at the level of data does not correspond to a
unique minimum at the level of PDFs: the datapoints are measurements of the
hadronic cross section $\sigma$ Eq.~(\ref{eq:fact}), which only
indirectly depends on the PDFs $f_i$. There is then a population of
PDFs which lead to the same optimal fit because of  the need  to
effectively interpolate between datapoints (``interpolation
uncertainty''). Namely, even though at the data level there is a
unique best fit, this does not correspond to a unique best-fit set of
underlying PDFs.

At level~1 the datapoints are fluctuated about their true values, so
the best-fit value of
figure of merit on datapoints  is now of order of $\chi^2\sim 1$
per datapoint. The uncertainty is correspondingly increased because
now there may be several PDF configurations which all lead to values
of the figure of merit of the same order, possibly corresponding to
different underlying functional forms for the PDFs (``functional
uncertainty''). In other words, now the prediction is no longer
uniquely determined even at the data level. Finally, at level~2,
corresponding to a realistic situation, the data themselves fluctuate
about the true value thereby inducing a ``data uncertainty'' on the
PDFs.

Figure~\ref{fig:level0} shows that for the gluon in the data region
these three components of the uncertainty are roughly of similar size.
Note that, if a fixed functional form was fitted to the data by
least-squares, both the level~0 and level~1 uncertainties would
necessarily vanish. Hence, to the extent that the final  level~2
uncertainty is faithful, a methodology based on a fixed functional
form, for which level~0 and level~1 uncertainties vanish, necessarily leads
to uncertainty underestimation.

\begin{figure}
    \center
    \includegraphics[width=.48\textwidth]{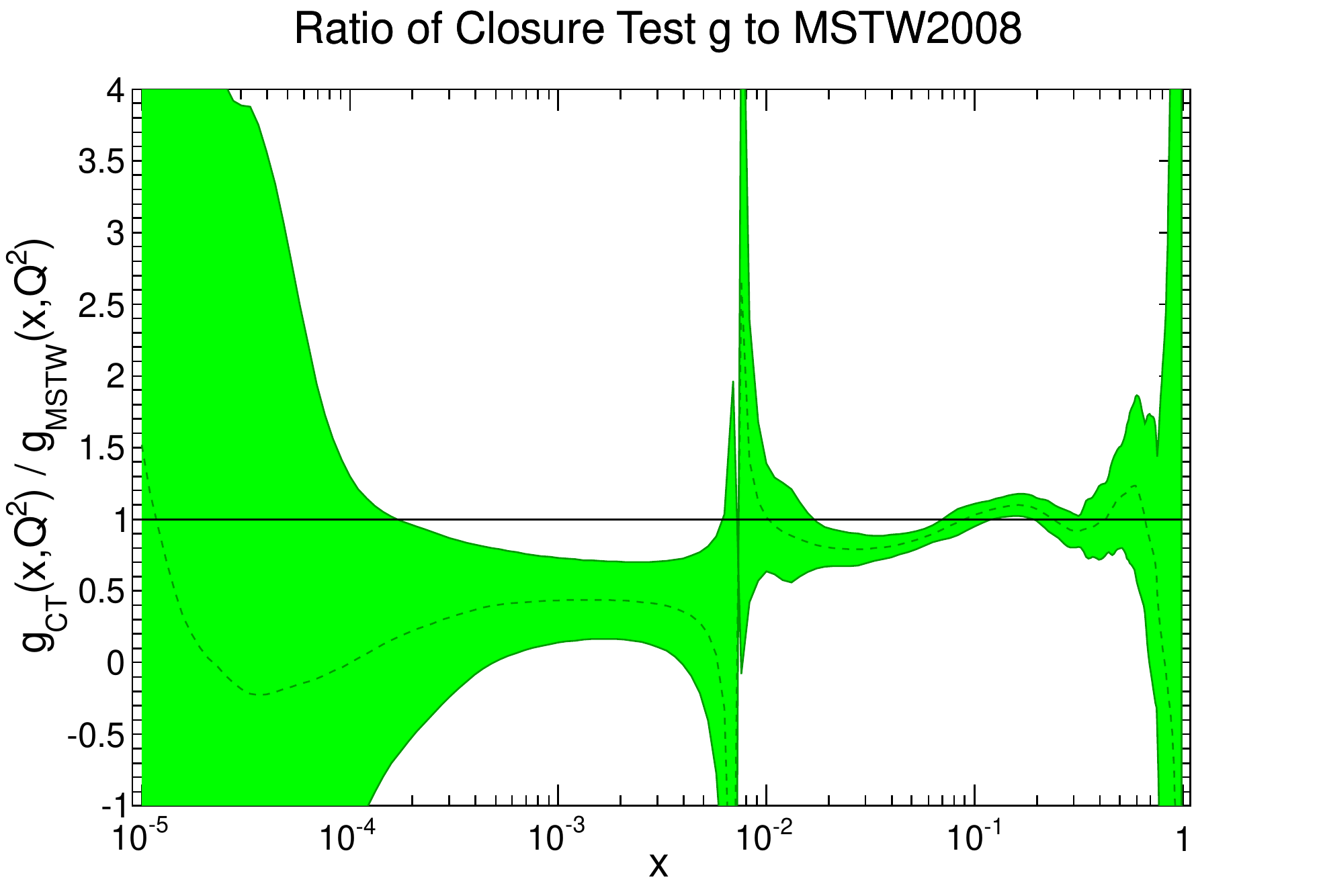}
    \includegraphics[width=.48\textwidth]{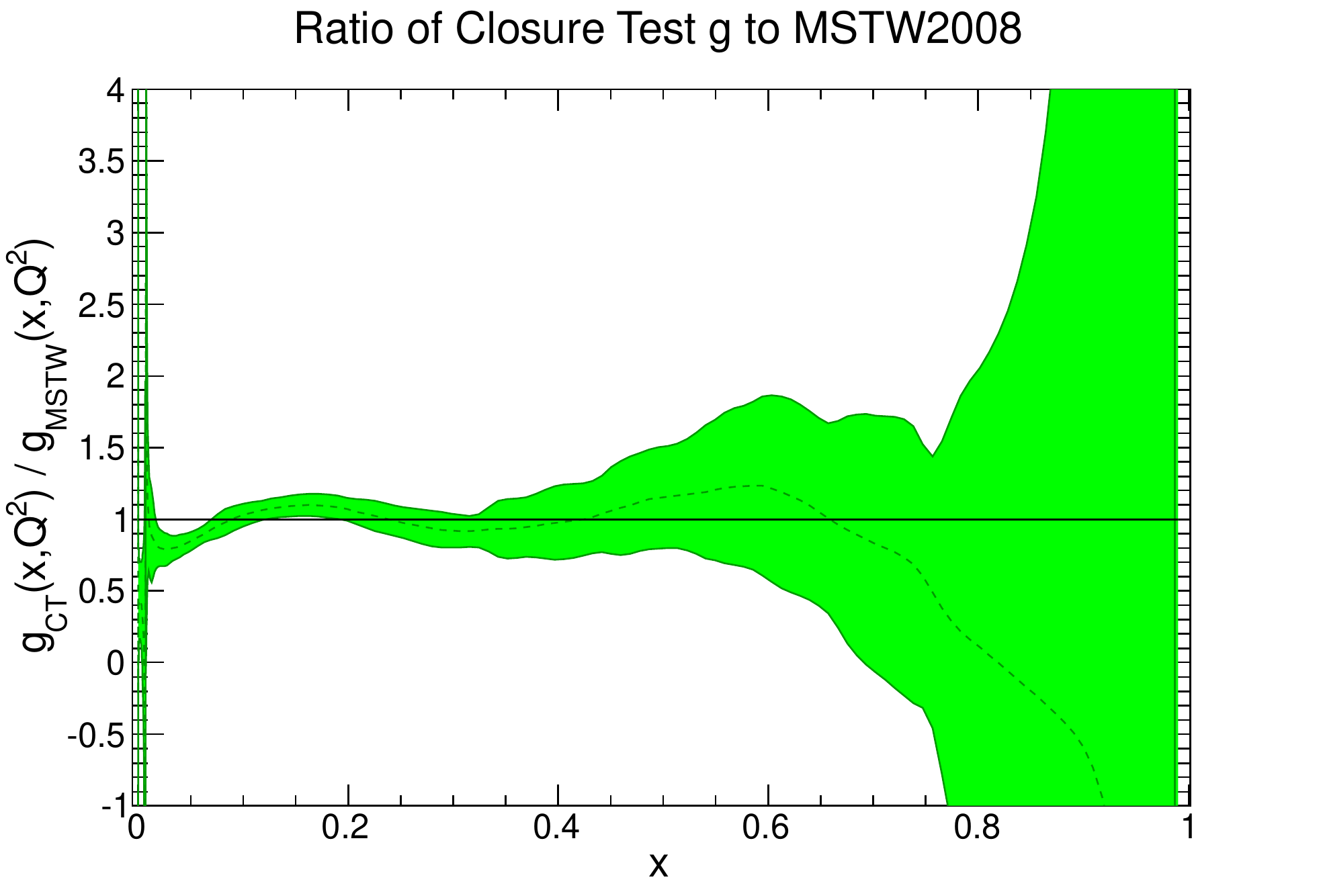}
    \caption{The best fit gluon compared to the underlying truth,
      shown vs. $x$ at the PDF
      parametrization scale  on a
      logarithmic (left) and linear (right) scale. The green band
      is the one-$\sigma$ uncertainty and the result is shown as a
      ratio to the underlying truth.}
    \label{fig:level2}
\end{figure}
This begs the question of checking whether indeed the level~2
uncertainties, namely, the uncertainties found with standard NNPDF
methodology are faithful. A first qualitative check can be done by
simply comparing the final result to the underlying truth, which in a
closure test is known. This is done for the gluon in
Figure~\ref{fig:level2}. It is clear that the result appears to be
broadly consistent: the truth is mostly within the one-$\sigma$ band,
though not always, which is as it should be, given that the one-$\sigma$ band
is supposed to be a 68\% confidence level. Note, however, that  PDF values
at neighboring points in $x$ are highly correlated: this is already true at the
level of single replicas, but even more for the final PDF, obtained
averaging over replicas, and it is of course as it should be -- after
all, if we were able to compute the PDF from first principles, it
would be given by a unique functional form, most likely infinitely
differentiable in the $0<x<1$ physical range. Hence, a confidence
level cannot be computed by simply counting how many point in $x$
space fall within  the one-$\sigma$ band.

Rather, a quantitative check that the confidence
level is correctly determined requires repeating the
whole procedure several times. Namely, we need to check that if
we regenerate a set of
(level~1) experimental values, and then refit them, in $68\%$ of
cases for each PDF at each point $f_i(x)$ the true value falls
within the one-$\sigma$ uncertainty. More in general, the validation
of the PDF determination requires first, computing PDFs and
uncertainties from a given set of level~2 data, so the PDF and
uncertainty are obtained by taking mean and covariance over replicas.
Next, repeating the
determination for different sets of level~2 data obtained from
different primary level~1 data: for each fit one will obtain a
different best-fit PDF set and corresponding uncertainties.
Finally, computing the distribution of
best-fit PDFs about the true value, and comparing this actual
distribution of results about the truth with their nominal uncertainty.

In practice, the procedure is  quite
costly as it requires producing a large enough number of
fits that confidence levels can be reliably computed,
each containing a large enough set of PDF replicas that the PDF
uncertainty can be reliably determined: for example, 100 sets of 100 PDF
replicas each. In Ref.~\cite{Ball:2014uwa} this was done by
introducing two approximations. First, the distribution of averages of
level~2 replicas, each from a  different set of level~1
data, was  approximated with  the
distribution of fits of a single replica to unfluctuated level~1 data.
Second,
the uncertainty was assumed to be stable between different fits and
was thus determined from a single 100-replica set to a particular set
of level~2 data. The validity of these approximations will be further
discussed in Sect.~\ref{sec:qct} below.
\begin{figure}
    \center
    \includegraphics[width=.65\textwidth]{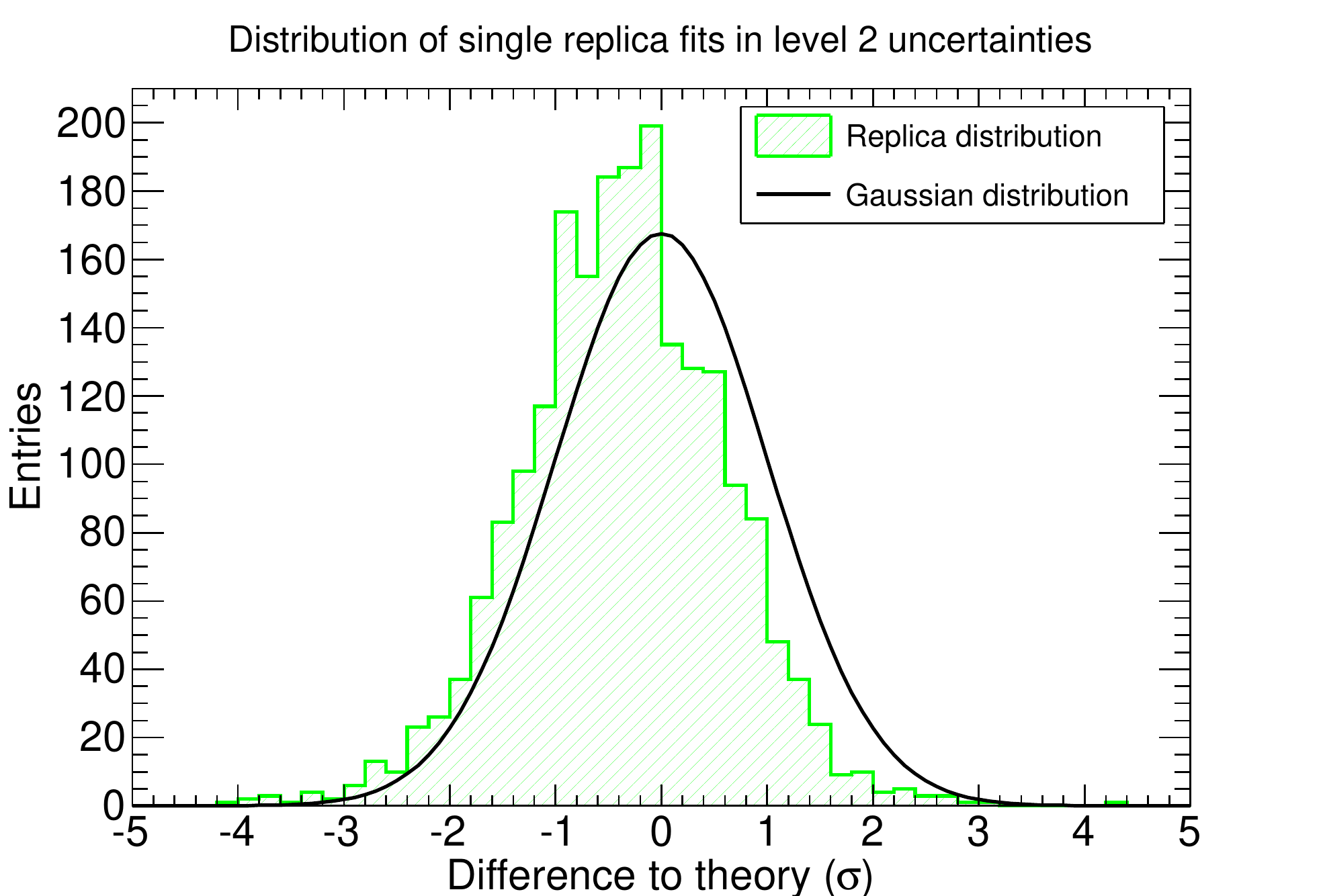}
    \caption{Distribution of deviation between the PDF and the
      underlying truth normalized to its nominal uncertainty, compared
      to an univariate Gaussian. Results are obtained sampling all
      fitted PDFs at three points in $x$.}
    \label{fig:histotest}
\end{figure}

This procedure was used in Ref.~\cite{Ball:2014uwa} to compute the
deviation of best-fit PDFs from the truth for all fitted PDFs evaluated at
three $x$ values: $x=0.05$,  $x=0.1$ and  $x=0.2$, and respective
uncertainties. The histogram  of normalized deviations is compared to
a univariate Gaussian in Figure~\ref{fig:histotest}. The deviation
between the predicted and observed probability distribution are small:
for instance, the one-$\sigma$
confidence level is 69.9\%, to be compared to the expected 68.3\%.
It is clear that
the validation is successful.

The availability of closure test data allows performing a variety of
further tests, all of which were done in Ref.~\cite{Ball:2014uwa}. On
the one hand, it is possible to compare to the truth various features
of the distribution of fitted PDFs, such as for example their
arc-lengths, or the behavior of their probability distribution upon
updating via Bayes' theorem. On the other hand, it is possible to test
the stability of results upon a number of variations of the methodology,
such as the choice of architecture of the neural nets, the choice of
GA and its parameters, the choice of PDF parametrization basis, the
parameters of the cross-validation. Indeed, as mentioned in
Section~\ref{sec:nn} it has been possible to check stability upon enlarging
the architecture of the neural net,  as mentioned ins
Section~\ref{sec:ga} the method was used in order to optimize the
parameters of the GA, and as mentioned above, it has been used to
check the stability with respect to
different choices of underlying truth.

\section{The future of PDFs in a deep learning framework}
\label{sec:PDFdl}

The AI-based approach to PDF determination described in
Section~\ref{sec:statart} largely eliminates potential sources of bias,
specifically those related to the choice of a functional form, as discussed in
Section~\ref{sec:pdfsai}, thanks to  the universal nature of neural
networks~\cite{Cybenko1989}.  However,
neural networks themselves are not unique, and the algorithms used for
their training even less so. The methodology discussed in
Section~\ref{sec:statart} has been developed over the years through a
long series of improvements,  as described in
Sections~\ref{sec:nn}-\ref{sec:ga}. These were based on trial and
error, and on the experience accumulated in solving a problem of
increasing complexity. The human intervention involved in these choices
might in turn be a source
of bias. A way of checking whether this is the case, and then improving on the
current methodology, is through hyperoptimization, namely, automatic
optimization of the methodology itself. This  goal was recently
accomplished, but it    required as a
prerequisite a redesign of the NNPDF codebase, and specifically the
replacement of the GA with deterministic minimization. Here we will
discuss first, this code redesign, next the hyperoptimization
procedure, then quality control, which plays a role analogous to
cross-validation but now at the hyperoptimization level, and finally,
the set of validation tests that ensure the reliability of the final
hyperoptimized methodology.

\subsection{A new approach based on deterministic minimization}
\label{sec:newdet}

The  NNPDF methodology presented in Section~\ref{sec:statart} was
implemented by the NNPDF collaboration as an
in-house software framework relying on few external libraries. There are two
major drawbacks of such an approach.
First, the in-house implementation greatly complicates the study of novel
architectures and the introduction of the  modern machine learning techniques
developed during the last decade.
Second, the computational performance of GA minimization algorithms
is a significant limitation, and it drastically reduces the possibility of
performing hyperparameter scans systematically.

In order to overcome these problems the code has been redesigned
using an object-oriented approach that provides the required functionality to modify and study
each aspect of the methodology separately, and a  regression model has
been implemented from scratch in a modular object
oriented approach based on external
libraries. Keras~\cite{chollet2015keras} and
TensorFlow~\cite{tensorflow2015:whitepaper} have been chosen as
back-ends for neural network and optimization algorithms. This code design
provides an abstract interface for the implementation of other machine learning
oriented technologies, that simplifies maintainability and opens the possibility
to new physics studies.

The new framework implements gradient descent (GD) methods to replace
the previously used GA described  in Section~\ref{sec:ga}. Thanks to
state-of-the art tools, this change reduces
the computing cost of a fit while achieving  similar or better
goodness-of-fit. The GD methods produce more stable fits than their GA
counterparts, and, thanks to the back-ends, the computation of the gradient of the
loss function is efficient even when including the convolution with the
FastKernel tables discussed in Section~\ref{sec:nn}. Given the
possibility of performing hyperoptimization scans, there is no longer
a risk of ending up in architecture-dependent local minima.

In terms of neural networks, the new code uses just one single
densely connected network as opposed to a separate network for each flavor. As
previously done, we fix the first layer to split the input $x$ into the pair
$(x, \log(x))$. We also fix 8 output nodes (one per flavor) with linear
activation functions. Connecting all different PDFs we can directly study
cross-correlation between the different PDFs not captured by the previous
methodology.

As we change both the optimizer and the architecture of the network,
the optimal setup must be re-tuned from scratch. To this purpose,  we
have implemented the hyperopt
library~\cite{Bergstra:2013}, which allow us to systematically scan over many
different combinations of hyperparameters finding the optimal configuration for
the neural network. Therefore, the neural network architecture no longer has
the form shown in Fig.~\ref{fig:arch}: first, rather than a neural net
per PDF, there is now a single
neural net with as many outputs as are the
independent PDFs, and second, the architecture (number of intermediate
layers and number of nodes per layer) is now hyperoptimized,
rather than being fixed.

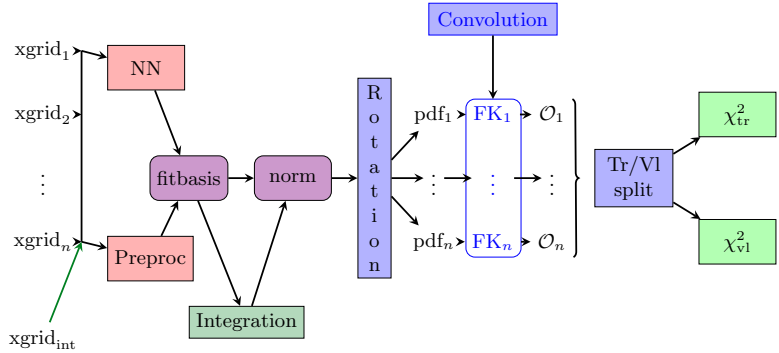
\begin{figure}[tb]
    \centering
    \resizebox{0.9\textwidth}{!}{%
    \begin{tikzpicture}[node distance = 1.0cm] \small
        \node (x1) {$\text{xgrid}_{1}$};
        \node[below of = x1] (x2) {$\text{xgrid}_{2}$};
        \node[below of = x2] (xd) {\vdots};
        \node[below of = xd] (xn) {$\text{xgrid}_{n}$};
        \node[below = 1cm of xn] (xint) {$\text{xgrid}_{\text{int}}$};

        \node[fitted, right = 1.0cm of x1.south, minimum width=1.2cm, minimum height=0.7cm]
                (pdf) {NN};
        \node[fitted, right = 1.0cm of xn.south, minimum width=1.2cm, minimum height=0.7cm]
                (preproc) {Preproc};

        \node[startstop, fill=violet!40, minimum width=1.2cm, minimum height=0.7cm, right = 1.5cm of xd.east]
                (fitbasis) {fitbasis};
        \node[startstop, fill=violet!40, minimum width=1.2cm, minimum height=0.7cm, right = 0.4cm of fitbasis]
                (normalizer) {norm};

        \node[fixed, minimum width=0.3cm, right = 0.4 of normalizer, text width=0.3cm]
                (rotation) {R o t a t i o n};

        \node[right = 0.5cm of rotation] (pd) {\vdots};
        \node[above of = pd] (p1){pdf$_{1}$};
        \node[below of = pd] (pn) {pdf$_{n}$};

        \node[blue,right = 0.6cm of pd] (fd) {\vdots};
        \node[blue,above of = fd] (f1) {FK$_{1}$};
        \node[blue,below of = fd] (fn) {FK$_{n}$};
        \node[procblue, minimum width = 2cm, minimum height=0.5cm, draw=blue, above = 1cm of f1] (convolution) {\color{blue}Convolution};
        \draw[draw=blue, rounded corners] (f1.north west) rectangle (fn.south east);

        \node[right = 0.6cm of fd] (od) {\vdots};
        \node[above of = od] (o1) {$\mathcal{O}_{1}$};
        \node[below of = od] (on) {$\mathcal{O}_{n}$};

        \node[fixed, right = 0.5cm of od, minimum width = 1.2cm, text width=1cm, minimum height=0.7cm]
                (trvl) {Tr/Vl split};
        \coordinate[right = 1cm of trvl] (cp);
        \node[n3py, above of = cp, minimum width = 1.2cm, minimum height=0.7cm]
                (chi2tr) {$\chi^{2}_\text{tr}$};
        \node[n3py, below of = cp, minimum width = 1.2cm, minimum height=0.7cm]
                (chi2vl) {$\chi^{2}_\text{vl}$};

            \node[procblue, fill=darkgreen!30, minimum width=1.0cm, minimum height=0.5cm] at ($(preproc) + (1.5, -1.0)$)
                (integration) {Integration};

        \draw[arrow] (pdf) -- (fitbasis);
        \draw[arrow] (preproc) -- (fitbasis);
        \draw[arrow] (fitbasis) -- (normalizer);
        \draw[arrow] (normalizer) -- (rotation);
        \draw[arrow] (trvl) -- (chi2tr);
        \draw[arrow] (trvl) -- (chi2vl);
        \draw[arrow] (integration) -- (normalizer);
        \draw[arrow] (fitbasis) -- (integration);
        \draw[arrow] (rotation) -- (p1);
        \draw[arrow] (rotation) -- (pd);
        \draw[arrow] (rotation) -- (pn);
        \draw[arrow] (p1) -- (f1);
        \draw[arrow] (pd) -- ($(fd) + (-0.33, 0.0)$);
        \draw[arrow] (pn) -- (fn);
        \draw[arrow] (f1) -- (o1);
        \draw[arrow] ($(fd)+(0.33,0.0)$) -- (od);
        \draw[arrow] (fn) -- (on);
        \draw[arrow] (convolution) -- (f1.north);

        \draw[decorate, decoration={brace}, thick] (o1.north east) -- (on.south east);


        \coordinate (a1) at ($(x1) + (0.6, 0.0)$);
        \draw[arrow] (x1) -- (a1);
        \draw[arrow] let
                \p1 = (a1), \p2 = (x2) in
                (x2) -- (\x1, \y2);
        \draw[arrow] let
                \p1 = (a1), \p2 = (xn) in
                (xn) -- (\x1, \y2);

        \draw[thick] let
                \p1 = (a1), \p2 = (xn) in
                (a1) -- (\x1, \y2);

        \draw[arrow] (a1) -- (pdf);
        \draw[arrow] let
                \p1 = (a1), \p2 = (xn) in
                (\x1, \y2) -- (preproc);

        \draw[arrow, darkgreen] let
                \p1 = (a1), \p2 = (xn) in
                (xint) -- (\x1, \y2);
    \end{tikzpicture}
    } \caption{Diagrammatic view of the \texttt{n3fit} code
    (from Ref.~\cite{Carrazza:2019mzf}){}.}
    \label{fig:n3fit}
\end{figure}

In Fig.~\ref{fig:n3fit} we show a graphical representation of the full new
methodology which will be referred to as \texttt{n3fit} in the sequel. The
$\text{xgrid}_{1}\dots \text{xgrid}_{n}$ are vectors containing the $x$-inputs
of the neural network for each of the datasets entering the fit. These values of
$x$ are used to compute both the value of the neural network and the
preprocessing factor, thus determining the unnormalized PDF. The normalization
constants $A_{i}$ (see Eq.~\eqref{eq:PDFdefinition}) are computed at
every step of the fitting using the
$\text{xgrid}_\text{int}$ points.
Recall from Section~\ref{sec:nn} that the PDFs are parametrized in a
basis of linear combinations
$\{g,\,\Sigma,\,V,\,V_3,\,V_8,\,T_3,\,T_8,\,c^+\}$: individual PDFs
for the quark flavors, antiflavors and the gluon,
$\{\bar{s}, \bar{u}, \bar{d}, g, d, u, s, c(\bar{c})\}$, are
obtained through a rotation. This procedure
concludes the necessary operations to compute the value of the PDF for any
flavor at the reference scale $Q_{0}$.

All PDF parameters are stored in two  blocks, the first named NN,
namely the neural network of Eq.~\eqref{eq:PDFdefinition}, and the
preprocessing $\alpha$ and $\beta$.
Given that each block is completely independent, we can swap them at any point,
allowing us to study how the different choices affect the quality of the fit.
All the hyperparameters of the framework are also abstracted and
exposed. This specifically allows us to
study several architectures hitherto unexplored in the context of PDF
determination.

As repeatedly discussed in Sections~\ref{sec:intro}-\ref{sec:statart},
the PDFs are not compared directly to the data, but rather,
predictions are obtained through  a
convolution over the neural networks. This, as mentioned
in Section \ref{sec:nn}, is performed through the FastKernel method, which
produces a set of observables $\mathcal{O}_{1}\dots\mathcal{O}_{n}$ from which
the $\chi^2$  Eq.~\eqref{eq:chi2} can be computed.
For this purpose, the first step is generation of
a rank-4 luminosity tensor
\begin{equation}
    \mathcal{L}_{i\alpha j \beta} = f_{i\alpha}f_{j\beta},
\end{equation}
where $(i,j)$ are flavor indices while $(\alpha, \beta)$ label the index on the
respective $x$ grids. Typical grids have of order of a hundred points
in $x$ for each PDF, spaced linearly in $x$ at large $x>0.1$, and
logarithmically at small $x$; the grids are benchmarked and optimized
in order to guarantee better than percent accuracy with high
computational efficiency~\cite{Ball:2010de,Bertone:2016lga,Ball:2017nwa}.
The physical observable, e.g. an inclusive
cross-section or differential distribution, is then computed by contracting the
luminosity tensor with the rank-5 FastKernel table for each separate dataset,
\begin{equation}
    \mathcal{O}_{n} = {\rm FK}^{n}_{i\alpha j \beta} \mathcal{L}_{i\alpha j \beta}, \label{eq:convolution}
\end{equation}
where $n$ corresponds to the index of the experimental data point within the
dataset. This stage of the model is the most computationally intensive.

\begin{figure}[tb]
    \centering
    \resizebox{0.5\textwidth}{!}{%
    \begin{tikzpicture}[node distance = 1.2cm]
        \node (init) [startstop] {training step};
        \node (ccheck) [check, below of = init] {counter $>$ max};
        \node (pcheck) [check, below of = ccheck] {positivity $>$ threshold};
        \node (xcheck) [check, below of = pcheck] {$\chi^2_\text{val}$ $<$ best $\chi^2$};
        \node (reset) [startstop, text width = 3cm, below of = xcheck] {reset counter  best $\chi^2 = \chi^2_\text{val}$};Q

        \node (cplus) [startstop, left = 0.7cm of ccheck] {counter ++};
        \node (end) [startstop, right = 0.7cm of ccheck] {END};

        \coordinate[above of = cplus] (li);
        \coordinate[below of = cplus] (lp);
        \coordinate[below of = lp] (lx);
        \coordinate[below of = lx] (lr);

        \draw[arrow] (init) -- (ccheck);
        \draw[arrow] (ccheck) -- node[right] {No} (pcheck);
        \draw[arrow] (pcheck) -- node[right] {Yes}(xcheck);
        \draw[arrow] (xcheck) -- node[right] {Yes}(reset);

        \draw[arrow] (ccheck) --  node[above] {Yes} (end);

        \draw[arrow] (reset) -- (lr) -- (lx);
        \draw[arrow] (xcheck) -- node[above] {No} (lx) -- (lp);
        \draw[arrow] (pcheck) --  node[above] {No} (lp) -- (cplus);
        \draw[arrow] (cplus) -- (li) -- (init);
    \end{tikzpicture}
    }
    \caption{Flowchart describing the patience algorithm of the
      \texttt{n3fit} code (from Ref.~\cite{Carrazza:2019mzf}){}.}
    \label{fig:stopping}
\end{figure}
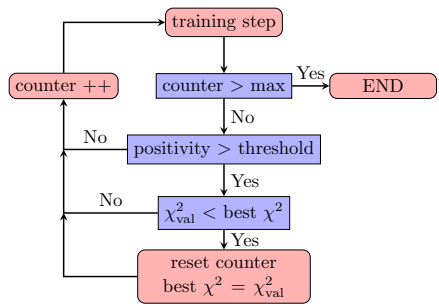

As discussed in Section~\ref{subsec:cv}, the optimal fit is determined
through cross-validation. The cross-validation split, which takes the output
and creates a mask for the training and validation sets, is introduced
as a final layer. As mentioned, the training
set is used for updating the parameters of the network during the fit while the
validation set is monitored during the fit and only used for early stopping
purposes. In Fig.~\ref{fig:stopping} we present a schematic view of the stopping
algorithm implemented in {\tt n3fit}. The training is performed until the
validation stops improving, from that point onward we enable a patience
algorithm which waits for a number of iterations before raising the stopping
action.
For post-processing purposes we only accept stopping points for which
the PDF
produces positive predictions for a subset of pseudo data which tests the
predictions for multiple processes in different kinematic ranges, see
Refs.~\cite{Ball:2014uwa,Ball:2017nwa} for further details.

The loss function Eq.~\eqref{eq:chi2} is minimized using gradient
descent.  Faster convergence and stability are found using
algorithms with adaptive moment, in which  the learning rate of the weights
is dynamically modified, such as Adadelta~\cite{ADADELTA},
Adam~\cite{Adam} and RMSprop~\cite{Tieleman2012}. These three optimizers
adopt similar gradient descent strategies,  but differ in the prescription for
weight update.

This approach has been applied to the  baseline
setup of the  NNPDF3.1 NNLO PDF determination~\cite{Ball:2017nwa}:
specifically, adopting  the
same dataset and cuts, together with the same fraction of validation data for
cross-validation, though now  the stopping criterion is different
(Fig.~\ref{fig:stopping}). This setup, henceforth referred to as
``global'',  includes all
datasets used in NNPDF3.1 NNLO, with 4285 data points.
We also studied a  reduced dataset which only includes data from
deep-inelastic scattering (DIS), which is computationally less intensive, in
particular because DIS is an electroproduction  process, so the
integral in Eq.~(\ref{eq:fact}) only involves a single PDF. This
setup, called ``DIS'', includes 3092 data points, and it facilitates
the process of benchmarking and validation, since it leads to
computationally  very light fits, which allow us to
extensively explore the parameter space.

\begin{table}
    \tbl{Comparison of the average computing resources consumed by the old and new methodologies for the DIS and Global setups.}
    {
    \begin{tabular}{c|ccc}
        \hline DIS fit & CPU h. & Mem. Usage (GB) & Good replicas \\ \hline
        \texttt{n3fit} (new)  & 0.2    & 2 & 95\%\\
        \texttt{nnfit} (old)  &  4    & 4 & 70\% \\
        \hline Global fit & CPU h. & Mem. Usage (GB) & Good replicas \\ \hline
        \texttt{n3fit} (new)  & 1.5    & 4 & 95\%\\
        \texttt{nnfit} (old)  & 30     & 5 & 70\% \\
        \hline
    \end{tabular}
    }
    \label{table:benchmark}
\end{table}

In summary, the new methodology considerably improves the
computational efficiency of PDF minimization, in particular because  GD
methods improve the stability of the fits,
producing fewer bad replicas  which need to be discarded, than theirs GA
counterparts. This  translates in a  much smaller computing
time. The old and new algorithms are compared in
Table~\ref{table:benchmark}:
we find a factor of 20 improvement with respect to
the old methodology and near to a factor of 1.5 in the percentage of accepted
replicas for a global fit setup.
In terms of memory, in the old methodology
usage is driven by the
APFEL~\cite{Bertone:2013vaa} code used in order to solve PDF evolution
equations, which does not depend on the set of experiments
being used. In the new code, evolution is never called during the fit
(it is pre-computed in the fktables and then the final PDFs are evolved
to all scales offline), so memory consumption is driven by the
TensorFlow optimization strategy which in the case of hadronic data requires the
implementation of Eq.~\eqref{eq:convolution} and its gradient. This difference
translates to an important decrease on the memory usage of \texttt{n3fit}.

\subsection{Optimized model selection}
\label{sec:hyperopt}

The main motivation for the development of the
new optimized code discussed in Section~\ref{sec:newdet} is the
possibility of performing systematic explorations of the methodology
through hyperoptimization. Firstly,
the new design of the {\tt n3fit} code exposes all parameters of the fit
including the neural network architecture. This is of key importance for a
proper hyperparameter scan where everything is potentially interconnected.
Furthermore, the new methodology has such a smaller impact on computing
resources that many more fits can be performed, with a difference by
several orders of magnitude:
for each fit using the old methodology hundreds of setups can now be tested.

The hyperparameter scan procedure has been implemented through  the hyperopt
framework~\cite{Bergstra:2013}, which systematically scans over a selection of
parameter using Bayesian optimization~\cite{Bergstra:2011:AHO:2986459.2986743},
and measures model performance to select the best architecture.
Table~\ref{table:scanparameters} displays an example of selection of
scan
parameters, subdivided into those which determine the
Neural Network architecture, and those which control the minimization.

\begin{table}
    \tbl{Parameters on which the hyperparameter scan is performed from~\cite{Carrazza:2019mzf}.}
    {
    \begin{tabular}{c c} \hline
        Neural Network & Fit options \\ \hline
        Number of layers & Optimizer \\
        Size of each layer & Initial learning rate \\
        Dropout & Maximum number of epochs \\
        Activation functions & Stopping  Patience \\
        Initialization functions & Positivity multiplier \\
        \hline
    \end{tabular}
    }
    \label{table:scanparameters}
\end{table}

Hyperparameter scans have been performed both in global and DIS
setups. The best model configuration has been searched for, using  as
input data
the original
experimental  values, rather than the data  replicas which are then
used for PDF determination (recall Section~\ref{sec:NNPDFappr}).
Optimization has been performed using
a combination of the best validation $\chi^2$ and stability of the fits:
specifically,  the architecture which produces the lowest validation
$\chi^2$ has been selected
after having trimmed  combinations which displayed unstable behavior.

\begin{figure}[t]
    \center
    \includegraphics[width=1.0\textwidth]{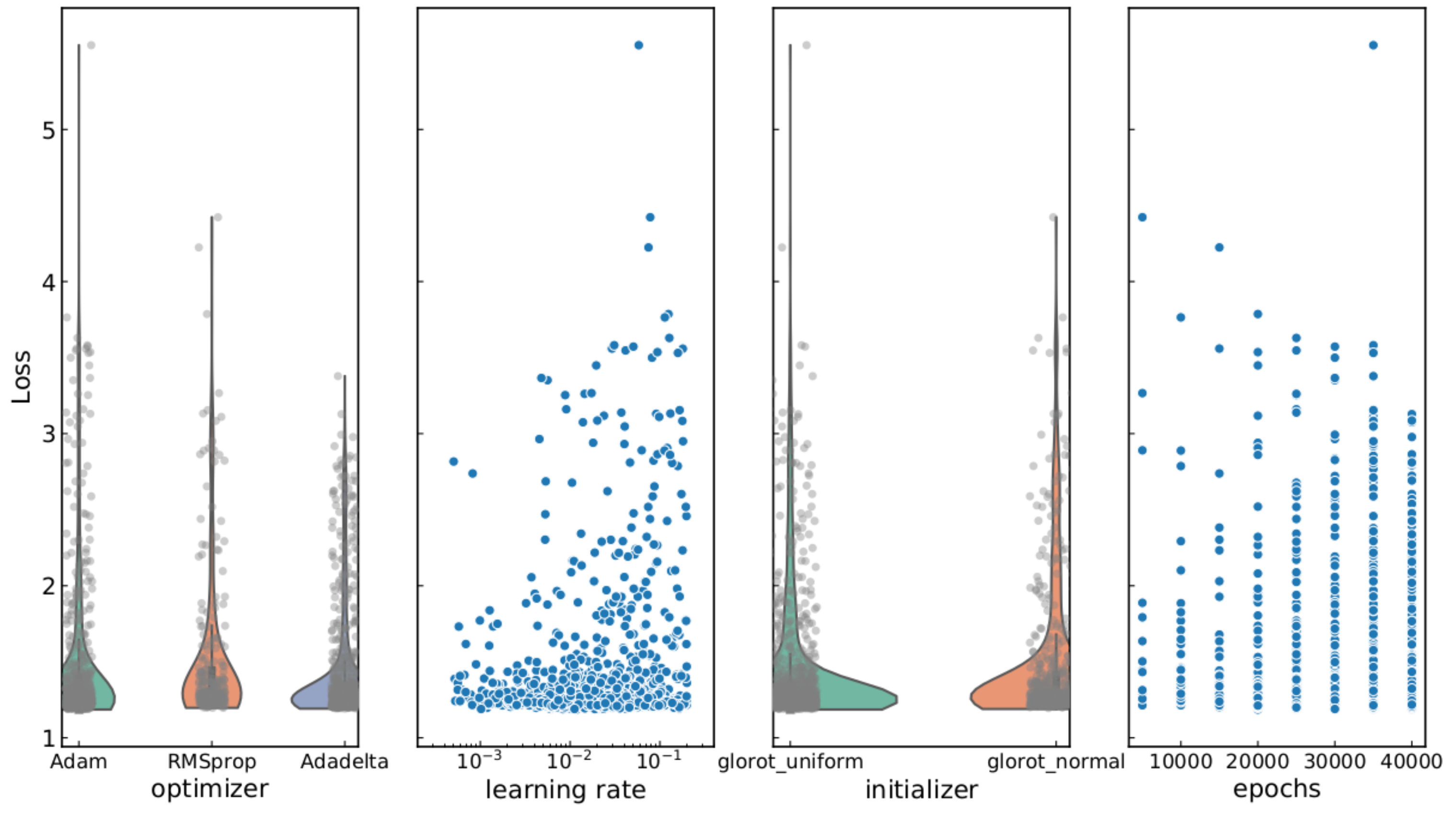}
    \includegraphics[width=1.0\textwidth]{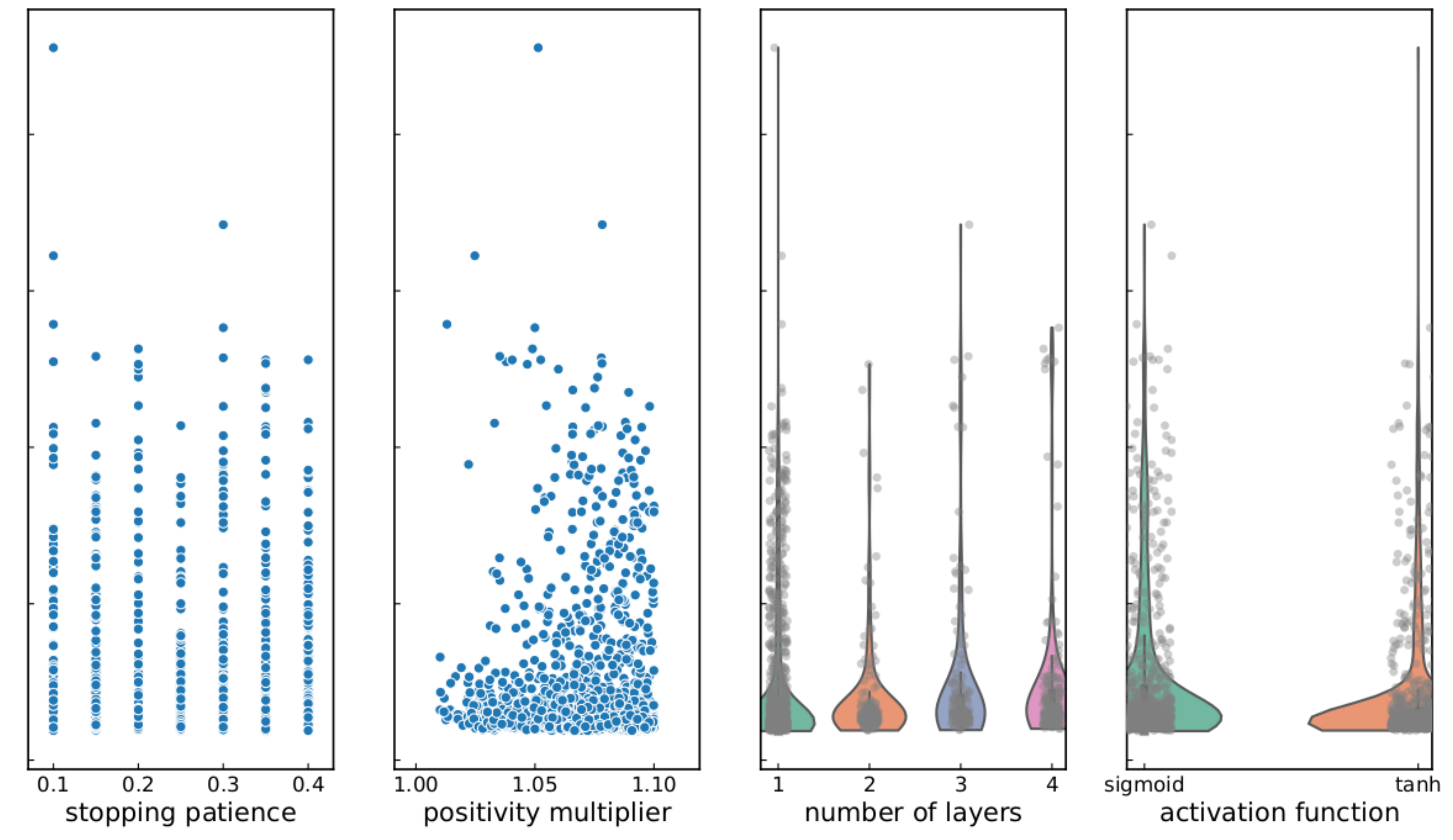}
    \caption{Graphical representation of a hyperparameter scan for a DIS only
      fit with 2000 trials (from Ref.~\cite{Carrazza:2019mzf}).
      The  loss function presented in the
    y-axis is  an average of the validation and testing $\chi^2$.
     The shape of the violin plots represent a visual aid on the
    behavior of the fit as a function of the free parameter. Fatter plots
    represent better stability,  i.e., configurations which are less likely
    to produce outliers.}
    \label{fig:hyperoptDIS}
\end{figure}

An example of scan for some of the
parameters shown in Table~\ref{table:scanparameters}, based the DIS setup,
is shown in Fig.~\ref{fig:hyperoptDIS}.
The results of this scan can be summarized as follows. The Adadelta
optimizer, for which no
learning rate is used, is found  to be more stable, and to systematically
produce better
results than RMSprop and Adam with a wide choice of learning rates. The
initializers, once unstable options such as a random uniform initialization have
been removed, seem to provide similar qualities with a slight preference for the
``glorot\_normal'' initialization procedure described
in Ref.~\cite{Glorot10understandingthe}{}.
Concerning the parameters related to stopping criteria, when the
number of epochs is very small the fit can be  unstable, however after
a certain threshold no big differences are observed. The stopping patience shows
a very similar pattern, stopping too early can be disadvantageous but stopping
too late does not seem to make a big difference. The positivity multiplier,
however, shows a clear preference for bigger values.
Finally, concerning the neural network architecture, a small
number of layers seems to produce slightly better absolute results, however, one
single hidden layer seems to lead to poor results. Concerning the
activation functions, the hyperbolic tangent seems to be
slightly preferred over the sigmoid. Once
an acceptable hyperparameter setup has been achieved, a final fine
tuning was performed, as some of the
choices could have been biased by a bad combination of the other parameters.

Clearly, the result of the hyperoptimization  depends on the
underlying dataset: for instance, we have verified that hyperoptimization on
a very large global dataset prefers a larger
architecture. Therefore, the reliability and stability of the
hyperoptimized methodology have to be checked a posteriori, as we will
discuss in Sect.~\ref{sec:qual}.

In summary, hyperoptimization has been implemented as a
semi-automatic methodology, that is capable of finding
the best hyperparameter combination as
the setup changes, e.g. with new experimental data, new algorithms or
technologies.

\begin{figure}[tb]
    \centering
    \includegraphics[width=0.65\textwidth]{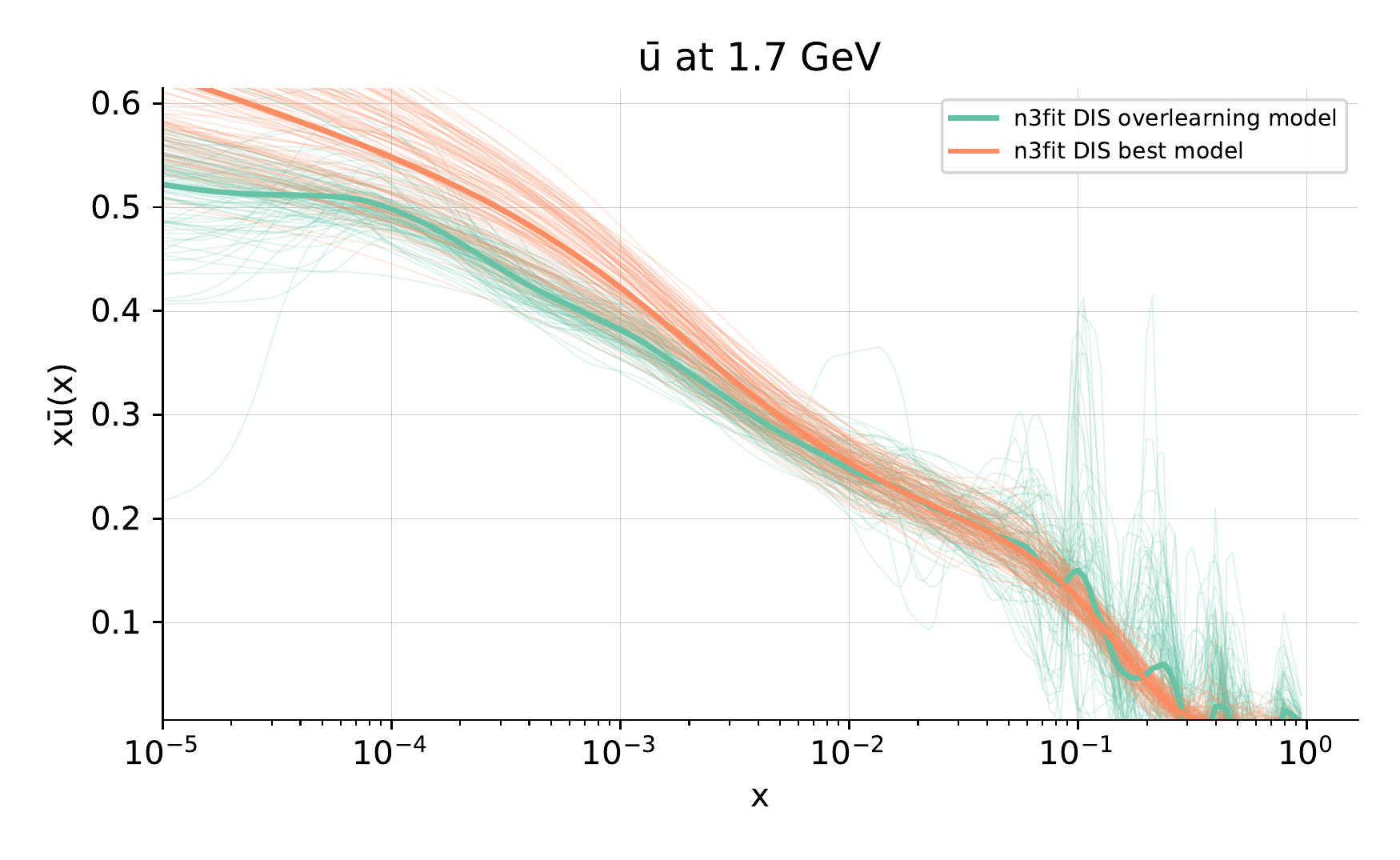}
    \caption{Comparison of replicas for the up quark PDF obtained by
      hyperoptimized  \texttt{n3fit} methodology without (green) and
      with (orange) quality control (from~\cite{Carrazza:2019mzf}).}
    \label{fig:overfitted}
\end{figure}
\subsection{Quality control}
\label{sec:newval}

The hyperoptimization presented in Sect.~\ref{sec:hyperopt} can be
viewed as a  meta-optimization in which the object of optimization is
the methodology. This immediately raises the issue of quality control.
In the fitting procedure, this is taken care by
cross-validation, in which quality control is provided by
the validation set.
A similar quality control is now needed at the hyperoptimization
level.

Indeed, if hyperoptimization is run by just optimizing on the
validation figure of merit, a typical result is shown in
Figure~\ref{fig:overfitted}, in which replicas for the up quark PDF
for a hyperoptimized DIS fit are shown. It is clear that an unstable
behavior is seen, characteristic of overtraining. This can also be
verified quantitatively:  for example the value of the training $\chi^2$
is much lower than that of the validation $\chi^2$.
This may appear to be surprising, given that the hyperoptimization is
performed on the validation $\chi^2$, while the training $\chi^2$ is
minimized in the fitting procedure. However, there
inevitably exist correlations between the training and validation sets,
for example through correlated theoretical and experimental
uncertainties. Due to these correlations, hyperoptimization without quality control
leads to overlearning.

\begin{figure}[tb]
    \centering
    \includegraphics[width=0.65\textwidth]{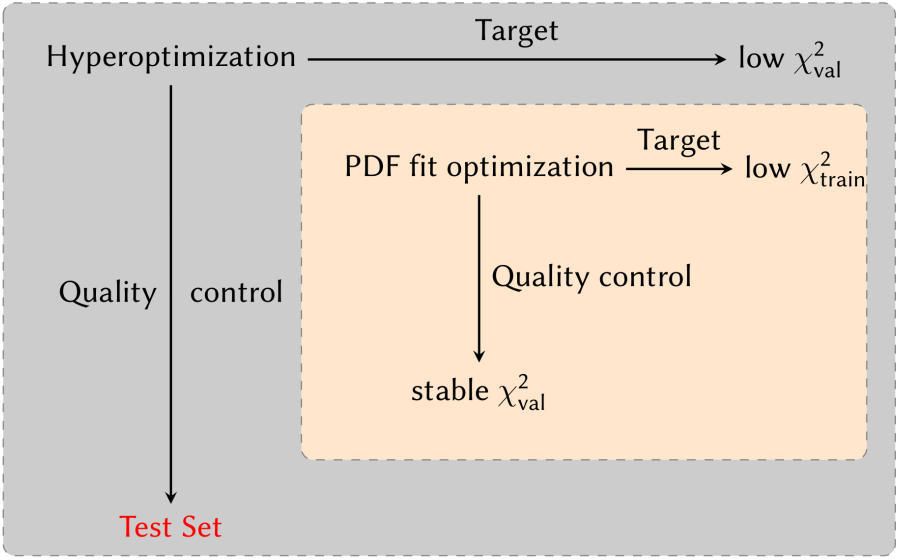}
    \caption{Schematic overview of the hyperparameter quality control methodology.}
    \label{fig:qa1}
\end{figure}
The problem can be solved by introducing a testing set, which tests
the generalization power of the model. The testing set is made out of
datasets which are uncorrelated to the training and validation
data, and none of which is  used in the fitting either for training or
validation. The test set plays the role of quality control for the
hyperoptimization, as schematically summarized in
Figure~\ref{fig:qa1}.

Defining the best appropriate test dataset for PDF fits is particularly
challenging due to the nature of the model regression through
convolutions. Indeed, the choice of prescription for the test set presents a
certain level of arbitrariness.
For a first exploration, the test set has been constructed by utilizing
datasets for which several experiments exist for the same process, and
picking the experiment with smallest kinematic range. The
corresponding data have been
removed from training and validation, and used as a test set.
A more refined option, which validates this first choice,
will be discussed in Section~\ref{sec:kfold} below.

\begin{table}[tb]
    \tbl{Best models found by our hyperparameter scan for the DIS and global setups using the new \texttt{n3fit} methodology.  }
    {
    \begin{tabular}{l|cc}
        \hline Parameter  & DIS only & Global \\ \hline
         Hidden layers & 2 & 3 \\
         Architecture & 35-25-8 & 50-35-25-8 \\
         Activation & tanh & sigmoid \\
         Initializer & glorot\_normal & glorot\_normal \\
         Dropout & 0.0 & 0.006 \\
         Optimizer & Adadelta & Adadelta \\
         Max epochs & 40000 & 50000 \\
         Stopping patience & 30\% & 30\% \\
         \hline
    \end{tabular}}
    \label{table:bestModel}
\end{table}

We have applied this procedure both to DIS and global fits. The
best models found in each case are compared in
Table~\ref{table:bestModel}. For the global setup deeper networks are allowed
without leading to overfitting. The hyperbolic tangent and the sigmoid functions
are found to perform similarly. The initializer of the weights of the network,
however, carries some importance for the stability of the fits, with
preference for the
Glorot normal initialization method~\cite{Glorot10understandingthe,glorot:2010}
as implemented in Keras.
Furthermore, adding a small dropout rate~\cite{DBLP:journals/corr/abs-1207-0580} to the hidden layers in the global fit reduces the chance of overlearning introduced by the deeper network, thus achieving more stable results.
As expected, the bigger network shows a certain preference for greater
waiting times (which also increases the stopping patience as is set to
be a \% of the maximum number of epochs). In actual fact, the maximum
number of epochs is rarely reached and very few replicas are wasted.

\begin{table}[tb]
    \tbl{Comparison of the total $\chi^{2}$ of the fit for both a DIS
      only and global fits found using the previous NNPDF3.1 and the
      new {\tt n3fit} methodology. }
    {
    \begin{tabular}{c|cc}
        \hline  & DIS only & Global \\ \hline
        \texttt{n3fit} (new) & 1.10 & 1.15 \\
        NNPDF3.1 (old) & 1.13 & 1.16  \\
         \hline
    \end{tabular}}
    \label{table:chi2fits}
\end{table}

\begin{figure}[tb]
    \includegraphics[width=0.5\textwidth]{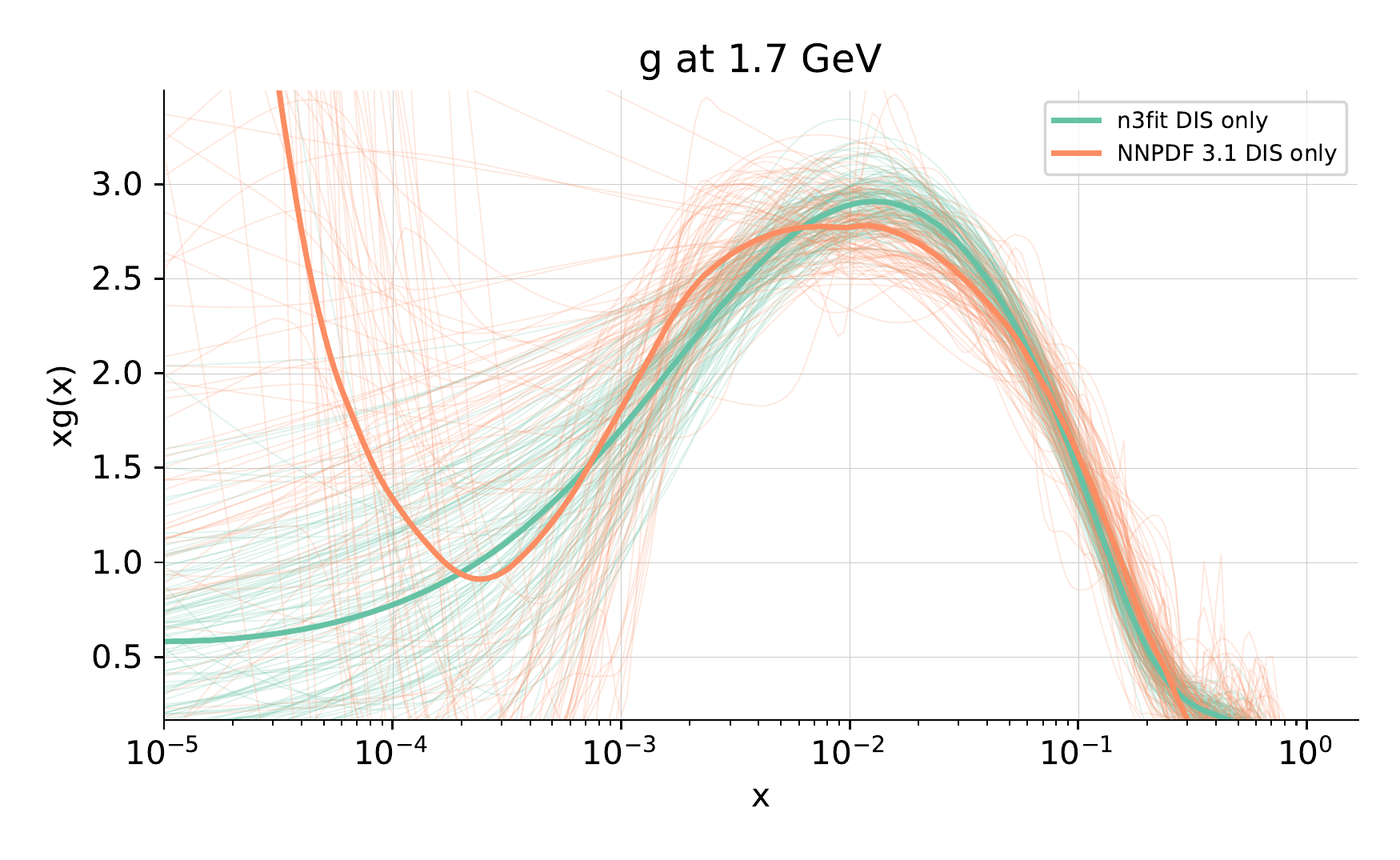}\includegraphics[width=0.5\textwidth]{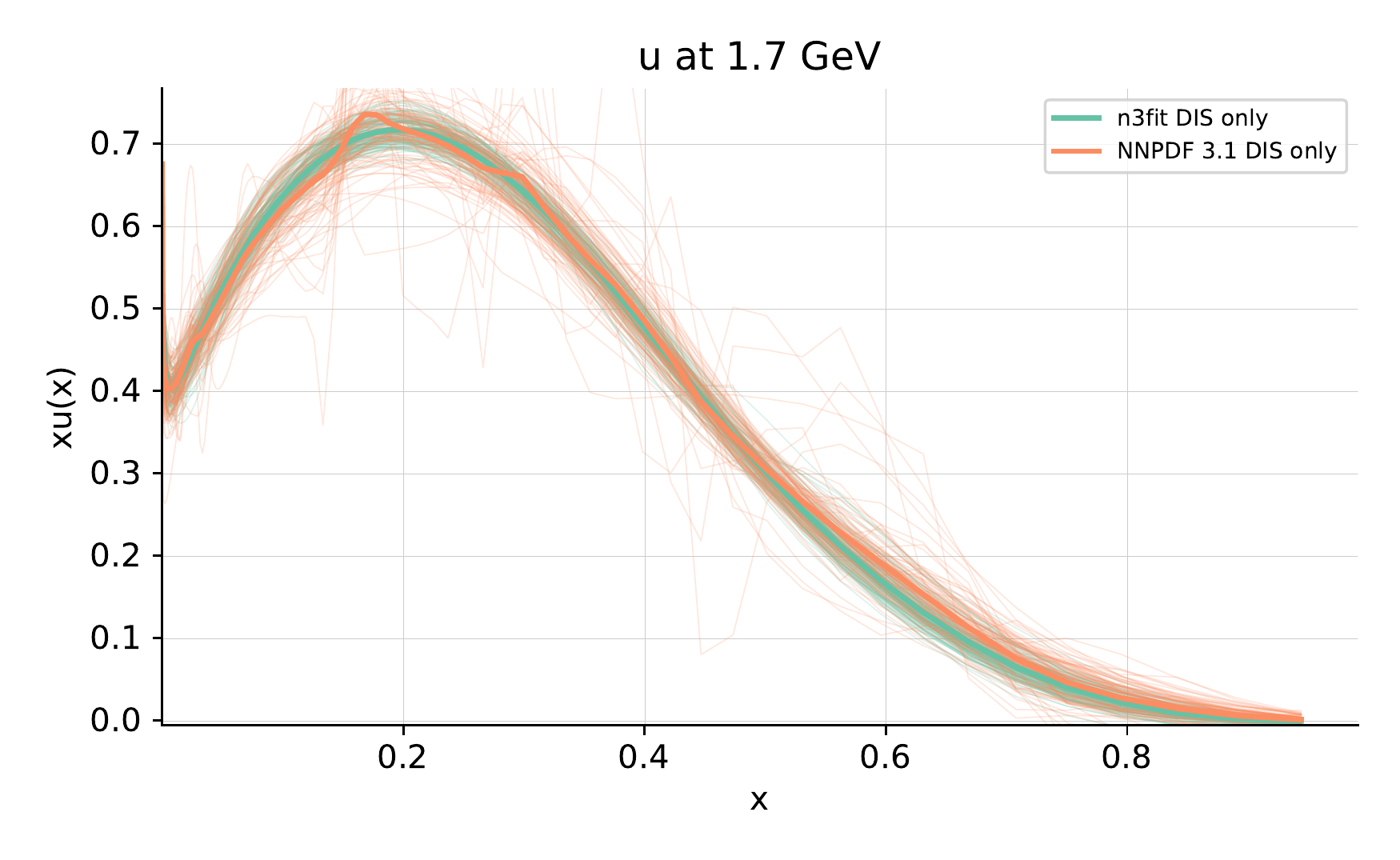}\\
    \includegraphics[width=0.5\textwidth]{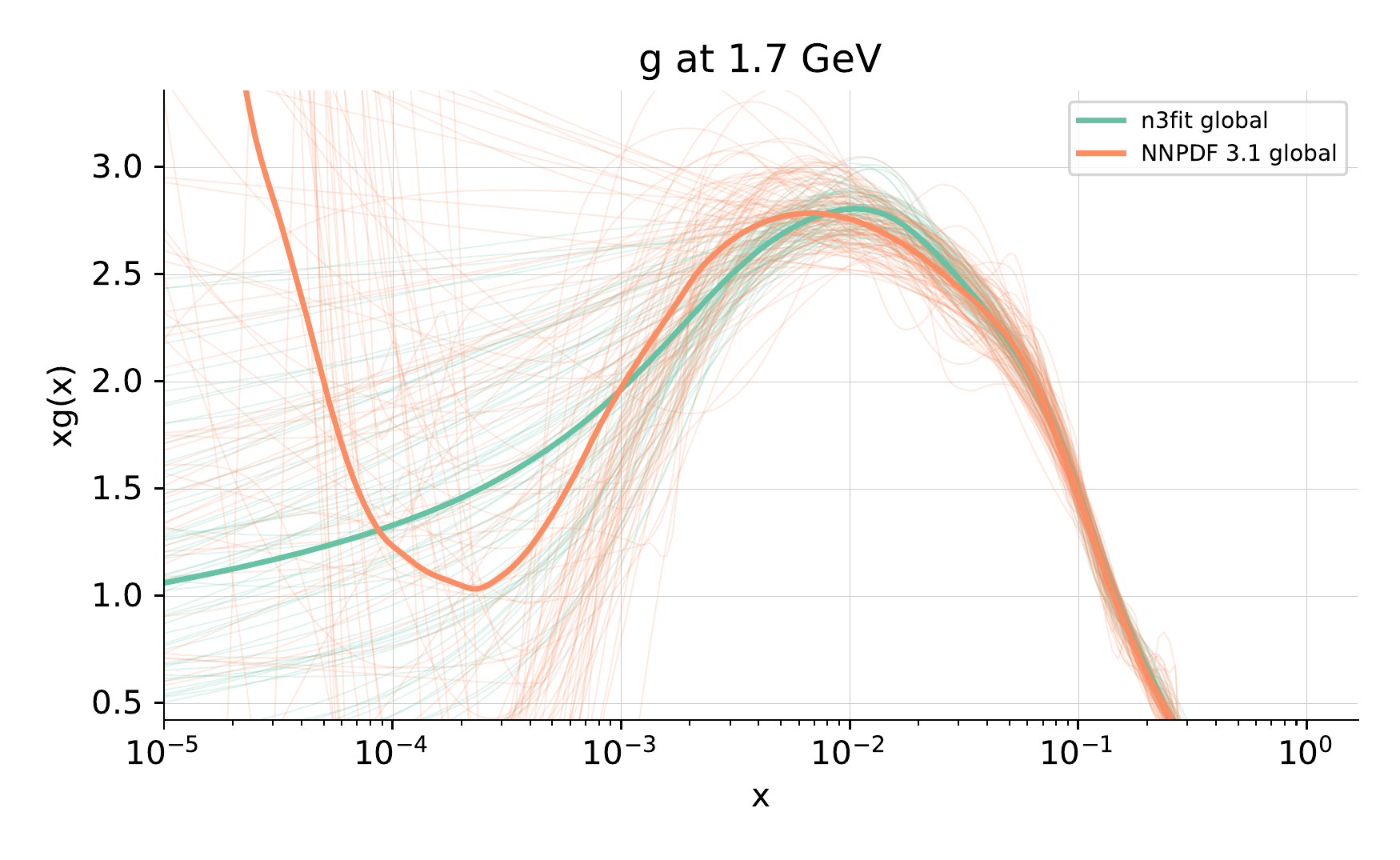}\includegraphics[width=0.5\textwidth]{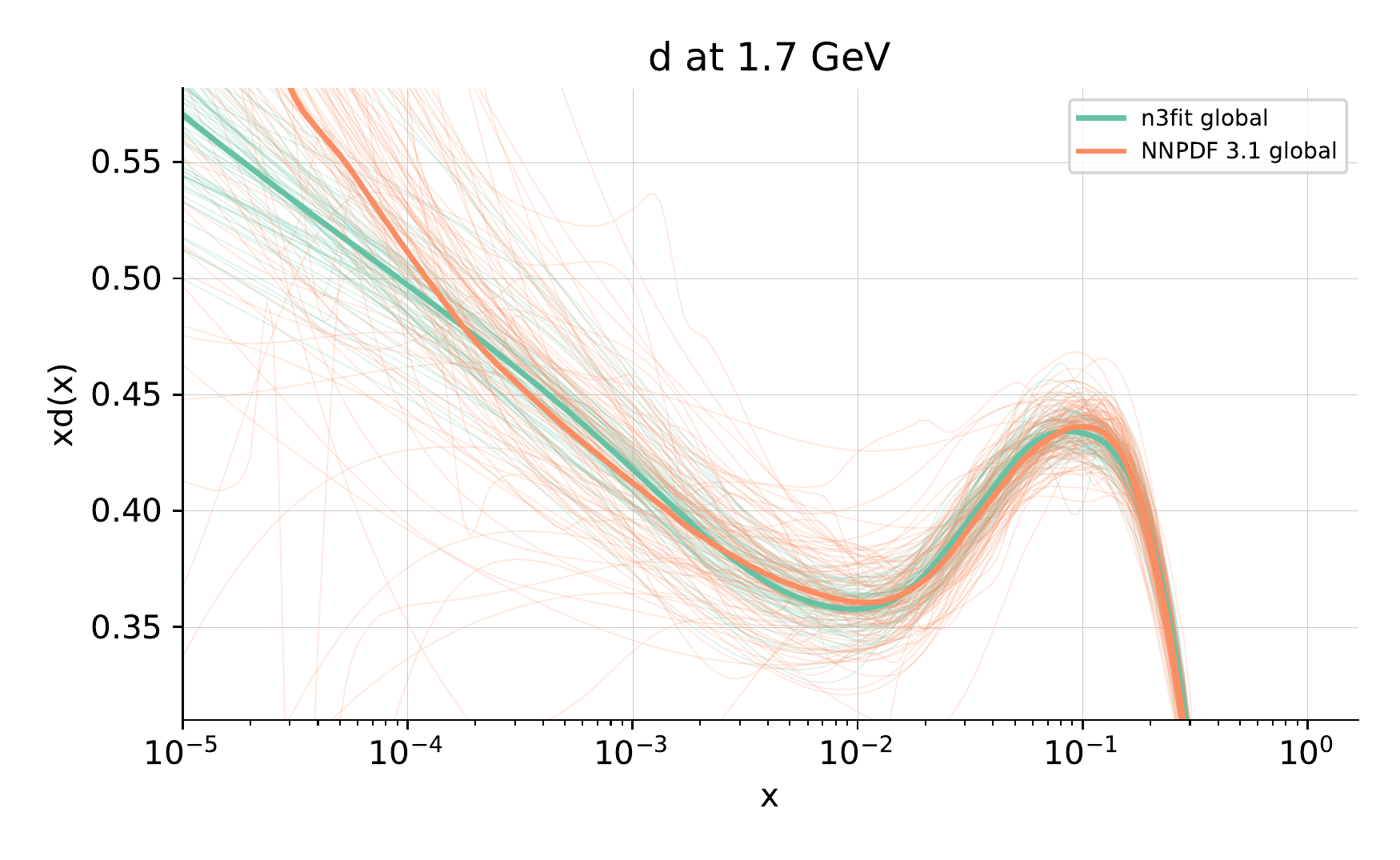}
    \caption{Comparison of PDFs found using the previous NNPDF3.1 and the
      new {\tt n3fit} methodology: for a DIS fit (top) the gluon
      (left) and up quark (right) are shown; for a global fit (bottom)
    the gluon (left) and down quark (right) are shown. (from~\cite{Carrazza:2019mzf}).}
    \label{fig:DISpdf}
\end{figure}

Turning now to  fit results,  despite the significant
difference in size and complexity of the dataset, the DIS and global
fits perform similarly in describing the experimental
data, as demonstrated by the  $\chi^{2}$ values presented in
Table~\ref{table:chi2fits}.
It is interesting to compare results to those obtained using the
previous NNPDF3.1 methodology. The total  $\chi^2$ values are compared in
Table~\ref{table:chi2fits}: even though the new
methodology leads to a slightly better fit, differences are small.
PDF replicas obtained with either methodology (for the gluon
and the up quark) are compared
Fig.~\ref{fig:DISpdf}, both for the DIS and global fits.
It is clear that the best-fit PDF, i.e. the
average over replicas, is not much affected by the change in
methodology
(though
somewhat smoother for {\tt nnfit}).

A significant difference however is seen at the
level of individual replicas: replicas found  with the new methodology
are rather
more stable, i.e. they fluctuate rather less. This leads to slightly
smaller uncertainties, and, more significantly,  with the
new methodology a
smaller number of replicas is necessary in order to arrive to a stable
average. The greater stability of the new methodology also leads to
somewhat smaller uncertainties in the far extrapolation, i.e. in
regions where there is no information and thus uncertainties are
large: this is seen in Fig.~\ref{fig:DISpdf} for the gluon
distribution for $x\lesssim 10^{-4}$. This raises the question of how
to reliably assess uncertainties in extrapolation: we will return to
this in Section~\ref{sec:future} below.

\begin{figure}[tb]
    \centering
    \includegraphics[width=0.49\textwidth]{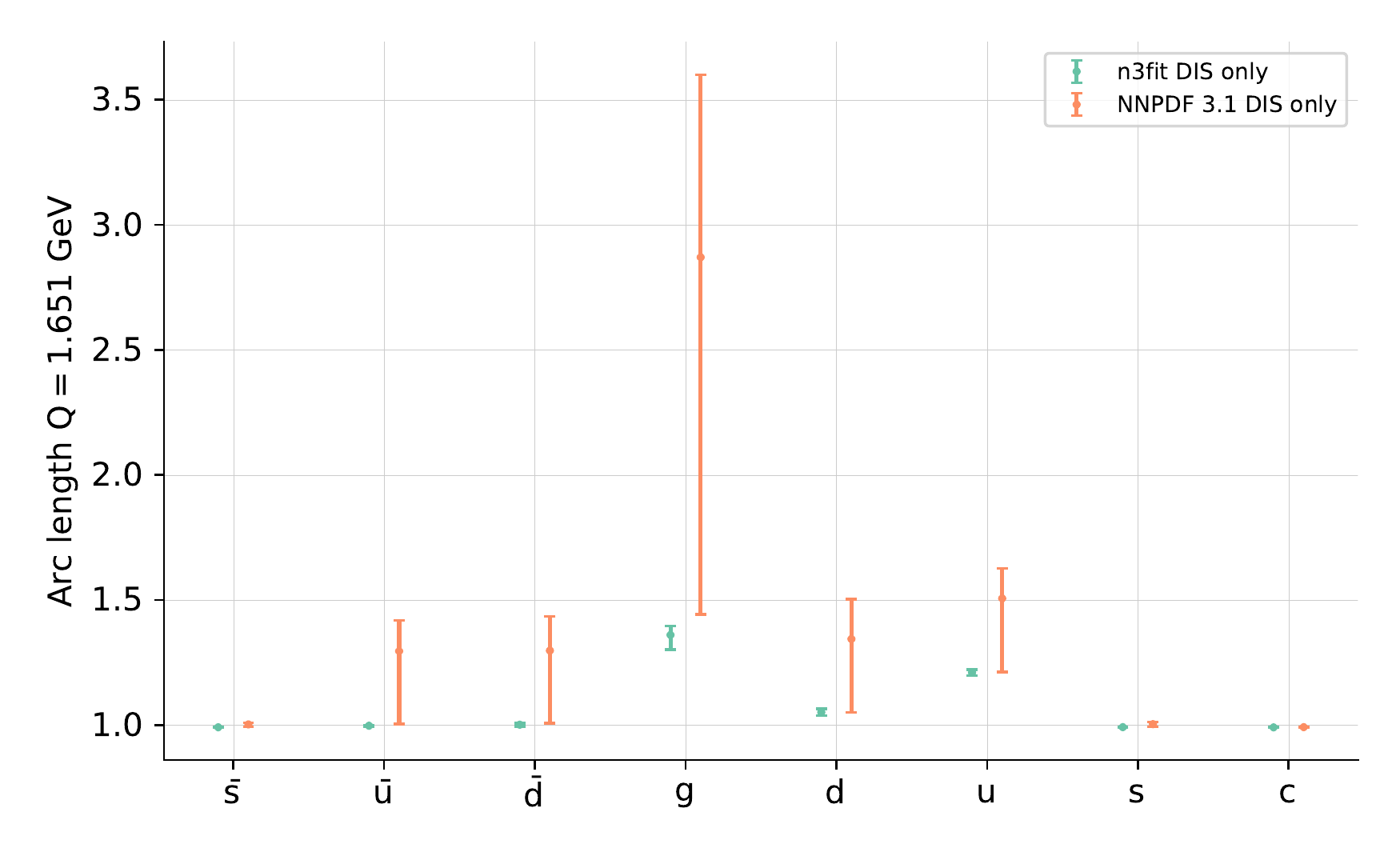}
    \includegraphics[width=0.49\textwidth]{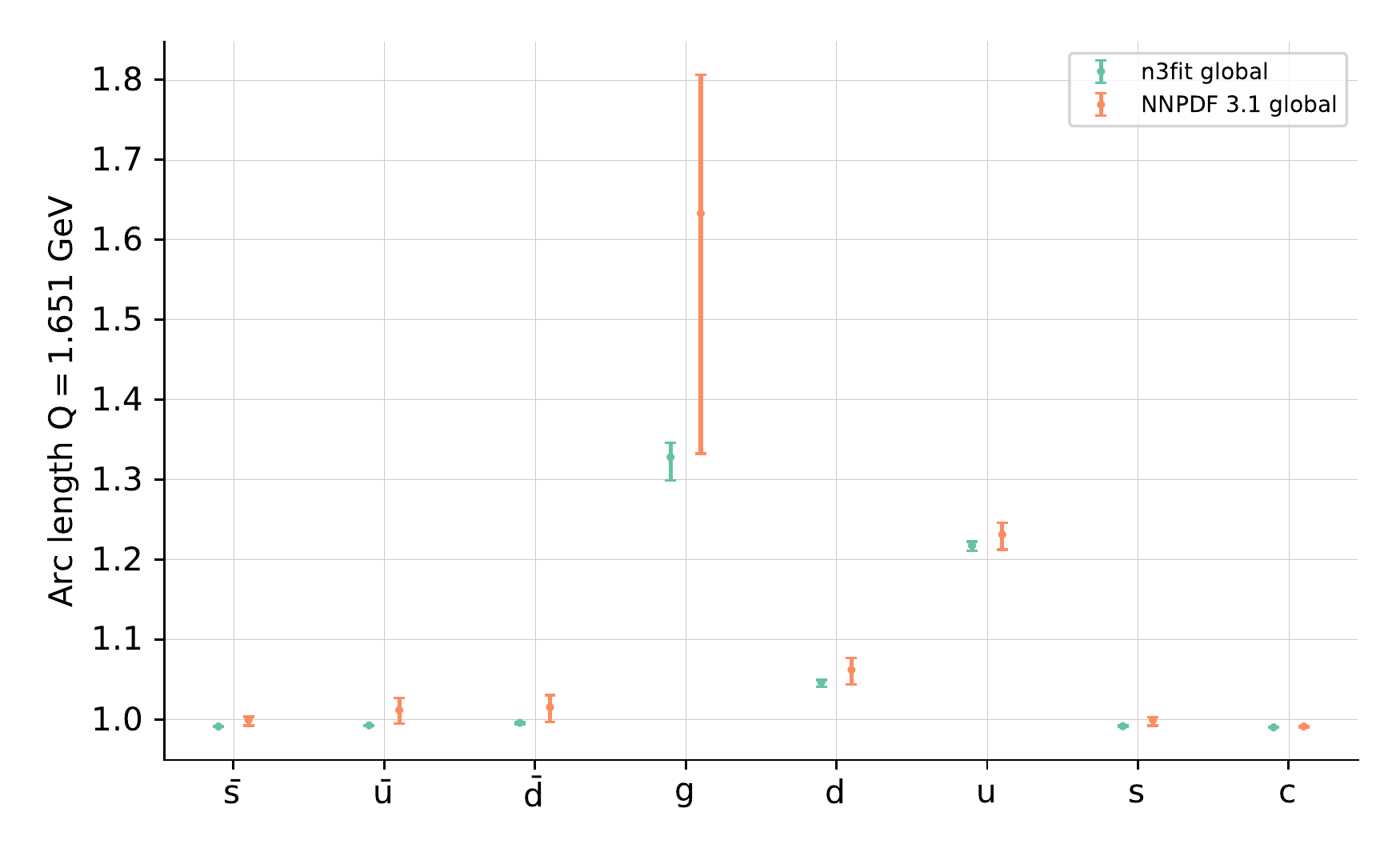}
    \caption{Comparison of PDF arc-lengths found using the previous
      NNPDF3.1 and the
      new {\tt n3fit} methodology in the DIS (left) and global (right)
      case. The mean and one-$\sigma$ interval computed from a set of
      PDF replicas for each PDF is shown.}
    \label{fig:DISArcLenghts}
\end{figure}
A particularly transparent way of seeing this greater stability
is to compare  PDF
arc-lengths. Because a PDF is a
function of $0<x<1$, one may define the length of the curve traced by
the PDF as $x$ varies in this interval. A very smooth PDF then has
smaller arc-length. In
Fig.~\ref{fig:DISArcLenghts} the mean and one-$\sigma$ values of
arclengths computed from  a set of replicas with the new and old
methodology are compared,
both for the DIS and global fits.
It is clear that, with the new methodology,
the arc-length  mean values are smaller, but especially
the fluctuation of  arc-length values between replicas is  much
smaller.

In summary we conclude that the new hyperoptimized {\tt n3fit}
methodology leads to results which are in broad agreement with the
current NNPDF3.1 methodology, thereby confirming that the latter is
faithful and unbiased, as expected based on the closure tests of
Section~\ref{sec:valid}. However, thanks to code redesign and
deterministic minimization it is possible to achieve greater
computational efficiency, and thanks to the hyperoptimization it
is possible to obtain, based on the same underlying datasets, more
stable results (i.e.,  a smaller number of replicas is sufficient to
achieve good accuracy) and somewhat smaller uncertainties. In short,
the new {\tt n3fit} methodology, while providing a validation of the
current NNPDF methodology,  displays greater computational efficiency, greater
stability  and  greater precision without loss of accuracy.
This in turn calls for more detailed validation and testing, as we now discuss.

\subsection{Validation and testing}
\label{sec:qual}

The {\tt n3fit} methodology motivates
and enables more detailed studies of fit quality.   It enables them because
  thanks to its much greater computational efficiency it is now
  possible to perform rather more detailed explorations than it was
  possible with the previous slower methodology.
 It motivates them, because
the goal of
  the new methodology is to allow for greater precision without loss
  of accuracy, namely, to extract more efficiently the information
  contained in a given dataset. It is then mandatory to make sure that
  no new sources of arbitrariness are introduced by the new
  methodology. Also, the new methodology is claimed to be more precise
  without loss of accuracy, i.e. to produce results which are more
  stable and have smaller uncertainty than the previous methodology
  given the same input. It is then crucial
  to
  perform validation tests which are  sufficiently detailed that the
  validity of this claim can be tested: in practice, this means tests
  that are sufficiently detailed that the
  two methodologies can be distinguished, and that impose more
  stringent requirements on the methodology itself.

  We will first discuss the new issue of robustness of the test-set
  methodology introduced in Section~\ref{sec:newval}, then turn to a
  more detailed set of closure tests, similar to those of Section~\ref{sec:valid}
 but now exploiting the new methodology, and finally discuss a new kind
 of test of the generalization power of the methodology: ``future testing''.

\subsubsection{Test-set stability}
\label{sec:kfold}

One new source of ambiguity in the {\tt n3fit} methodology
is the choice of an appropriate test set.
Indeed, the setup discussed in Section~\ref{sec:newval}  was based on
a particular choice of test set, but one would like
to avoid as much as possible this kind of potentially biased
subjective choice. Also, in that setup one has to discard some data
from the dataset used for fitting and only include them in the test
set. This contrasts with the desire to keep data in the
training set as much as possible, in order to exploit as much as
possible the (necessarily limited) dataset in order to determine the
wide variety of features of the underlying PDFs.

These goals can be achieved through a $k$-fold cross-validation.
 In this algorithm, data are subdivide into  $k$ partitions, each of
 which reproduces the broad features of the full dataset. Each of the
 partitions then plays in turn the role of the test set, by being
 excluded
 from the fit. A variety of figures of merit
can then be chosen for hyperparameter optimization, such as
the mean  value of the loss  over excluded
partitions, or the best worst value of the validation loss of the
excluded partition.

\begin{figure}[tb]
    \centering
    \includegraphics[width=0.5\textwidth]{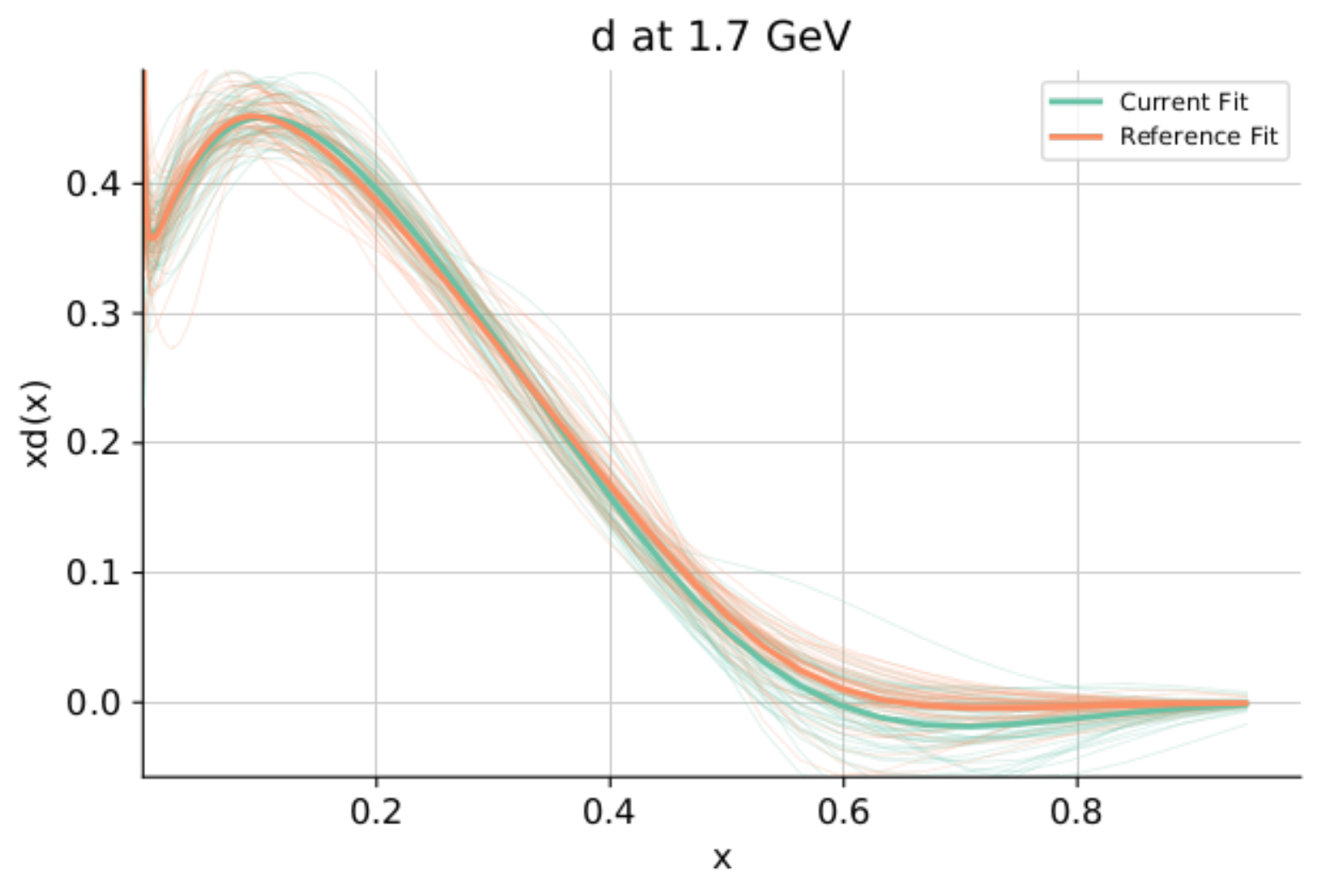}\includegraphics[width=0.5\textwidth]{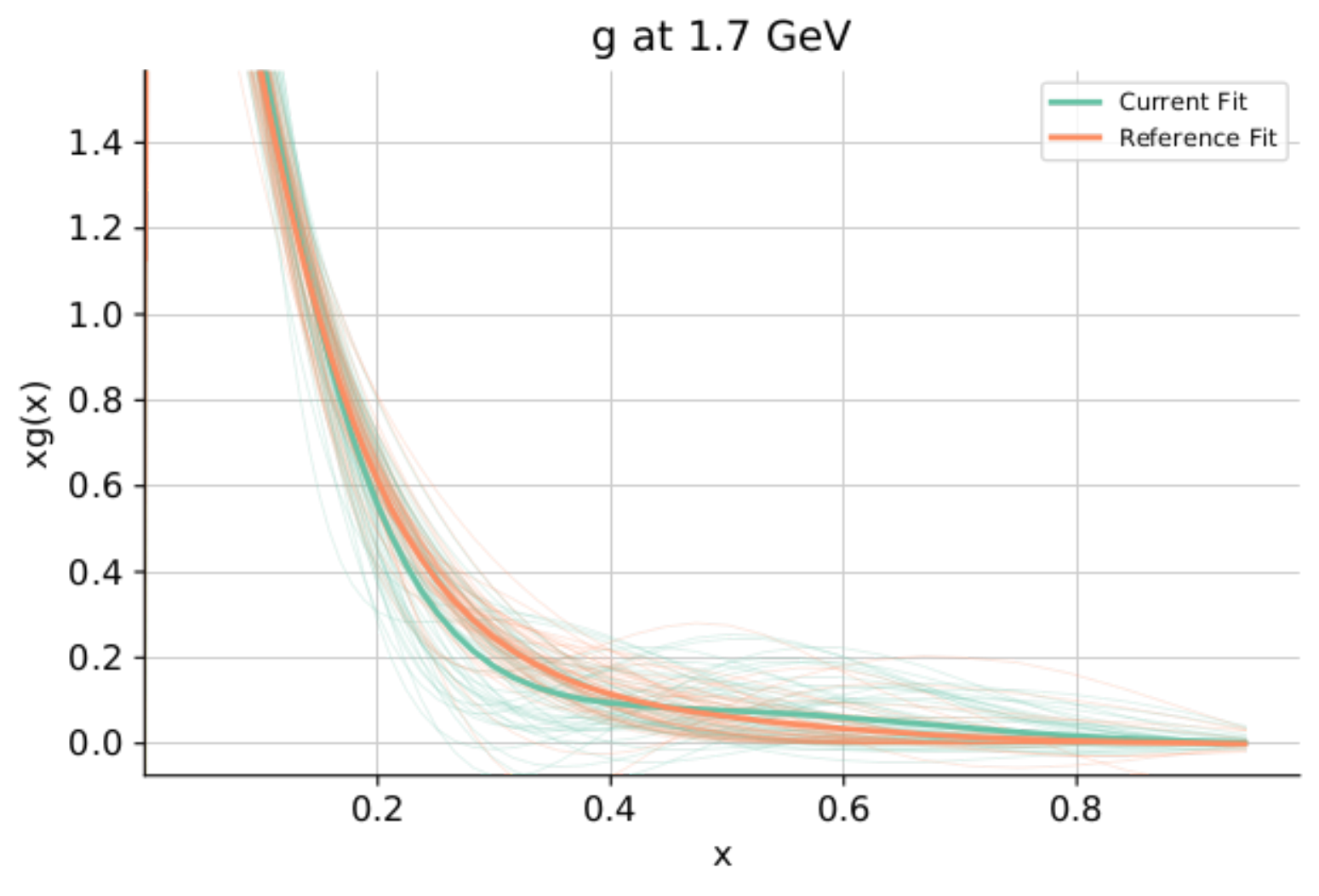}
    \caption{Comparison between the best models from $k$-fold cross-validation (green) and manual selection (red)\cite{NNPDF4}.}
    \label{fig:qa2}
\end{figure}

This $k$-folding procedure  has been implemented, and stability upon
different choices of hyperoptimization figure of merit has been
explicitly checked.
Results are shown in
Figure~\ref{fig:qa2}, where  the best PDF models estimated using
$k$-folding
are compared to those obtained through the simple test-set
procedure of Section~\ref{sec:newval}.
Similar results are found using either method. While confirming
the reliability of the manually selected method of
Section~\ref{sec:newval}, this allows us to replace it with the more
robust and unbiased $k$-folding method.

\subsubsection{Closure testing}
\label{sec:qct}

We now turn to closure testing, as presented in
Section~\ref{sec:valid} in the context of NNPDF3.0~\cite{Ball:2014uwa}.
We have applied the closure  testing methodology of
Section~\ref{sec:valid}, but now using
the {\tt n3fit} methodology and the more recent and wider
NNPDF3.1~\cite{Ball:2017nwa} dataset and
theory settings. Hence,  level~2 data are now in one-to-one
correspondence with data in the  NNPDF3.1 dataset, and, more
importantly, we can take advantage of the greater computational
efficiency of {\tt n3fit}.

  A first example of this
   is that it is now possible to perform confidence
  level tests based on actual full reruns. Indeed, recall from
  Section~\ref{sec:valid} that a computation of a closure test
  confidence level requires producing several independent fits, each
  with a sufficiently large number of replicas, so that the population
  of central values and uncertainties in each fit can be compared to
  an underlying truth. Thanks to the use of {\tt n3fit},
  it has now been possible
  to perform 30 different closure test level~2 fits, each with 40
  replicas~\cite{wilsondebbio}.
  Results are
  then further enhanced and stabilized by using  bootstrapping, i.e.,
  by drawing random subsets of fits and random subsets of replicas
  from each fit and computing the various estimators for the resample
  of fits and replicas. It has been possible to check in this way that
  results are essentially stable with at least 10 fits with at least
  25 replicas each, in that increasing the number of fits and replicas
  results are unchanged. All numbers quoted below refer to results
  obtained with the largest numbers of fits and replicas. The fact
  that such a relatively small
  number of replicas is sufficient to achieve stable result is a
  reflection of the greater stability of {\tt n3fit} replicas
  discussed in Section~\ref{sec:qual}.

  \begin{figure}
    \center
    \includegraphics[width=.65\textwidth]{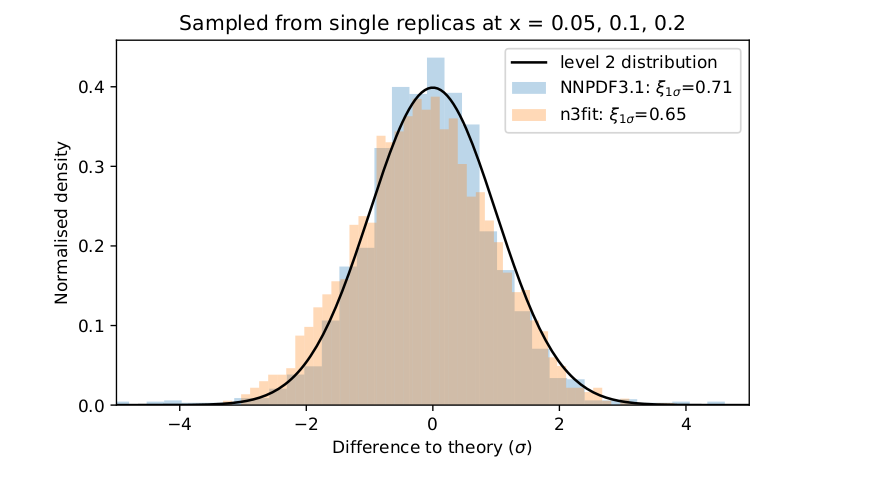}
    \caption{Same as Fig.~\ref{fig:histotest}, but now using NNPDF3.1
      data and methodology, and comparing results obtained using the
      approximate methodology of Section~\ref{sec:valid} (NNPDF3.1
      methodology) and the exact methodology ({\tt n3fit} methodology)\cite{wilsondebbio}.}
    \label{fig:histotest3}
  \end{figure}
  As a first test, we recompute the histogram of deviations of
  Figure~\ref{fig:histotest}, but now using NNPDF3.1 data. We
  can now compare the histogram actually computed using 30 fits
  with 40 replicas each, with the histogram approximately
  determined using a single 100 replica level~2 fit and 100
  single-replica level~1 fits, as it was done for
  Figure~\ref{fig:histotest} (labeled ``NNPDF3.1 methodology''). The
  result is shown in Figure~\ref{fig:histotest3}. It
  is clear that  the validation is successful also for the (rather
  wider) NNPDF3.1
  dataset: the one-$\sigma$ confidence level now equal to 65\%, and
  the mean of the histogram is now essentially unbiased, unlike
  in  Figure~\ref{fig:histotest} were a small bias was present.
  Also
  the approximate method used in
  Section~\ref{sec:valid} and Ref.~\cite{Ball:2014uwa} is reasonably
  accurate: specifically, the true value 65\% is reasonably well approximated
  by  the value 71\% found using the approximate method.

  We can now proceed to more detailed closure tests by computing
  confidence levels more
  extensively . A useful tool in this context is the
  bias-variance ratio. This, for Gaussian distributions,  contains
  exactly the same information as the one-$\sigma$ confidence level of
  predicted values
  with respect to the underlying truth considered in
  Section~\ref{sec:valid}.
  For uncorrelated data, the bias-variance ratio
  is defined as the mean square deviation of the
  prediction from the truth (bias), divided by the expected one-$\sigma$
  uncertainty (variance). The square-root of the bias-variance ratio
  \begin{equation}\label{eq:rbv}
    R_{bv}=\sqrt{
      \frac{1}{N_{\rm dat}}\sum_{i=1}^{N_{\rm dat}}\frac{(d_i-d_i^{(0)})^2}{\sigma_i^2}}
      \end{equation}
(where $d^i$, $\sigma_i$ and $d_i^{(0)}$ are respectively the
  prediction, uncertainty and true value for the $i-th$ datapoint)
  is
the ratio between observed and predicted
  uncertainties, and thus it should be equal to one for a
  perfect fit. The generalization to the correlated case is
  straightforwardly obtained by  expressing the numerator and
  denominator under the square root in Eq.~(\ref{eq:rbv})
  in terms of the covariance matrix.
  We have verified explicitly that the value of the
  one-$\sigma$ confidence level interval computed using the measured
  bias-variance ratio coincides with the measured confidence level,
  within statistical accuracy, so
  either can be equivalently used.

  We can now turn to more detailed comparisons.
  First, the comparison can be done for each
  PDF individually, rather than for all PDFs lumped together. Second,
  the comparison can also be done at the level of experimental data:
  namely, instead of determining the deviation between the fitted and
  true PDF we determine the deviation between the prediction obtained
  using the best-fit PDF and the true PDF for each of the datapoints
  in the NNPDF3.1 dataset.

  It should be noted that of course the
  predictions for individual datapoints are correlated due to the use
  of  common underlying PDFs, with
  correlations becoming very high for datapoints which are
  kinematically close, so that the integral Eq.~(\ref{eq:fact}) is
  almost the same. These correlations can  be simply determined  by
  computing the covariance matrix between all datapoints
  induced by the use of the underlying PDFs, which in turn is done by
  determining covariances over the PDF replica sample.
  Confidence levels are then determined
  along eigenvectors of this  covariance matrix, and can be compared
  to the bias-variance ratio, either by using its general form in the
  non-diagonal data basis, or
  equivalently, using  Eq.~(\ref{eq:rbv}) but with the sum running not on the
  original datapoints, but rather over the eigenvectors of the
  covariance matrix.

  Of course,
  the PDFs themselves are also correlated. The histograms in
  Figures~\ref{fig:histotest},\ref{fig:histotest3} were computed by
  sampling each PDF at three widely spaced points in $x$ so as to
  minimize this correlation, but of course computing a histogram of
  deviations with correlations neglected is still an
  approximation. When performing comparisons in PDF space we have now
  therefore also computed the covariance between PDFs over the replica
  sample, and determined confidence intervals  along its eigenvectors,
  and the corresponding bias-variance ratio values with correlations
  kept into account.

  A first comparison has been performed by computing the bias-variance
  ratio at the data level. This leads to an interesting result. Recall
  from Section~\ref{sec:valid} and Figure~\ref{fig:level012} that the
  total PDF uncertainty consists of three components of comparable
  side, the first of which is due to the need to interpolate between
  data. Clearly, this latter component is absent if one compares the
  prediction to the same data which have been used  to produce
  the PDF set. Indeed, we find that the square root of the
  bias-variance ratio computed for the NNPDF3.1 dataset (more than
  4000 datapoints) is
  $R_{bv}=0.74$. If we compute the same ratio for a new wide dataset including
  about 1300
  HERA, LHCB, ATLAS and CMS data not used in the fit we find that the
  value is $R_{bv}=0.9$. The difference between these two values can
 be understood as an indication of the fact that in the former case
 the bias does not include the level~1 uncertainty, while the variance
 (which should be used for new prediction) does. The value
 $R_{bv}=0.9$ means that PDF uncertainties on predictions
  are accurate to
  10\% (and somewhat overestimated).

  \begin{table}
    \tbl{The bias-variance ratio $R_{bv}$ Eq.~(\ref{eq:rbv} and the
  one-sigma confidence level for individual PDFs, computed using four
  points in $x$ space per PDF along eigenvectors of the covariance matrix\cite{wilsondebbio}.}
{\begin{tabular}{c|cc}
\hline
PDF    \qquad\qquad\qquad\qquad\qquad\qquad                       &  $R_{bv}$      &  one-$\sigma$ c.l.      \\
\hline
$\Sigma$  \qquad\qquad\qquad\qquad\qquad\qquad                         & 0.9    &   70\% \\
gluon  \qquad\qquad\qquad\qquad\qquad\qquad     & 0.9    &  69\%  \\
V   \qquad\qquad\qquad\qquad\qquad\qquad   & 1.0    &  66\%  \\
V3  \qquad\qquad\qquad\qquad\qquad\qquad  & 1.0   &   93\% \\
V8  \qquad\qquad\qquad\qquad\qquad\qquad  & 0.9    & 71\%   \\
T3  \qquad\qquad\qquad\qquad\qquad\qquad  & 0.6    & 89\%   \\
T8  \qquad\qquad\qquad\qquad\qquad\qquad  & 1.3    & 46\%   \\
\hline
total \qquad\qquad\qquad\qquad\qquad\qquad & 0.9 & 0.71 \\
\hline
\end{tabular}}
\label{tab:pdfbv}
\end{table}


  We next computed both the bias-variance ratio and the one-sigma confidence
  level at the PDF level. PDFs have been sampled at four points for
  each PDF, in a region in $x$ corresponding to the data region, and
  the covariance matrix has been subsequently diagonalized as
  discussed above. Results are shown in Table~\ref{tab:pdfbv} for
  individual PDF combinations. It is clear that, especially for the PDF
  combinations that are known with greater accuracy, such as the quark
  singlet $\Sigma$ and the gluon $g$, uncertainties are
  faithful: only the combination $T8$ which measures the total
  strangeness shows a certain amount  of uncertainty underestimation,
  by about 30\%.

\subsubsection{Chronological future tests}
\label{sec:future}

  \begin{figure}
    \center
    \includegraphics[width=.65\textwidth]{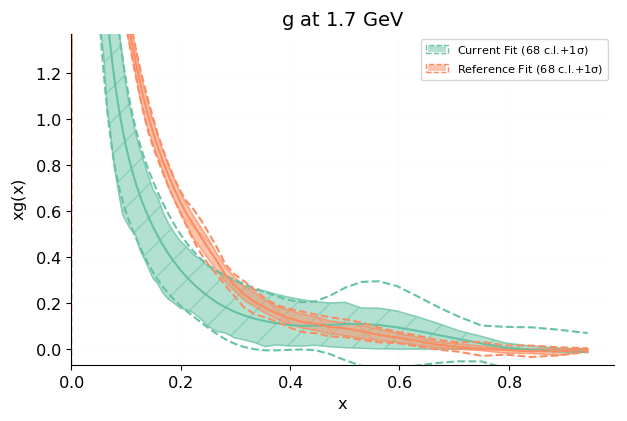}
    \caption{The gluon PDF determined used pre-HERA data (green)
      compared to the current best-fit (orange)\cite{NNPDF4}.}
    \label{fig:futureg}
  \end{figure}
The closure tests essentially verify the reliability of results in the
data region. A much more difficult task is to verify the power of
generalization of the methodology: namely, whether PDFs determined with
a subset of data are able to correctly predict the behavior of new
data, including those that extend the kinematic domain used for PDF
determination. In practice, this means testing whether PDF
uncertainties are reliable also in  regions in which they start
growing significantly because of lack of information.

This is done by ``chronological'' or ``future'' tests. Namely, we
consider an existing (or hypothetical) past dataset, we train PDFs
based on it, and we compare the best-fit results with later data which
extend the kinematic region. A first test of this kind has been
performed only including data which predated the HERA electron-proton
collider, and which thus approximately correspond to the information
on PDFs available around 1995. This is especially interesting since
it is well known (see e.g.~\cite{Tung:2004rw}) that the best-fit gluon
shape substantially changed after the advent of HERA data, as pre-HERA
data impose only very loose constraints on the gluon PDF.

\begin{figure}[tb]
    \centering
    \includegraphics[width=0.5\textwidth]{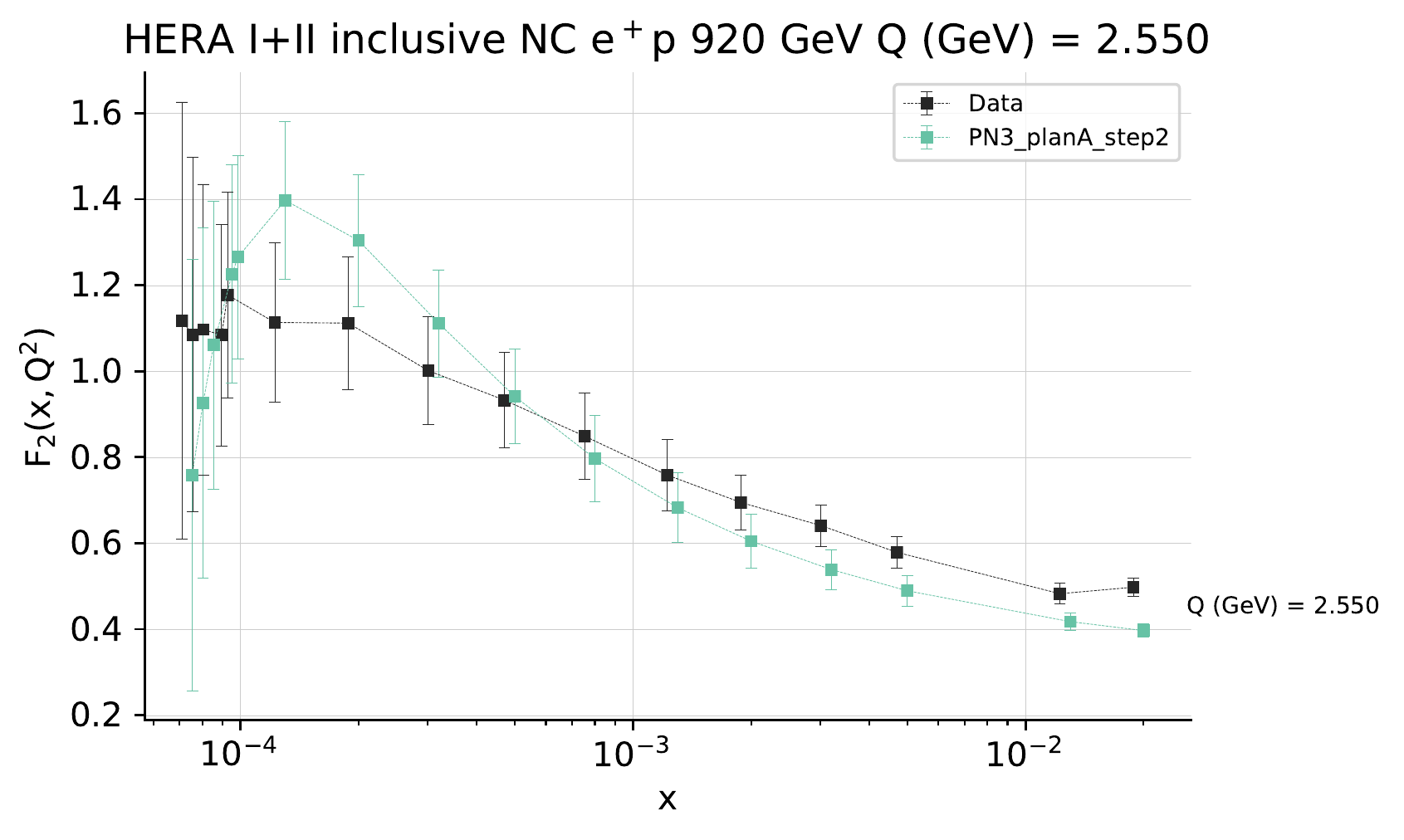}\includegraphics[width=0.5\textwidth]{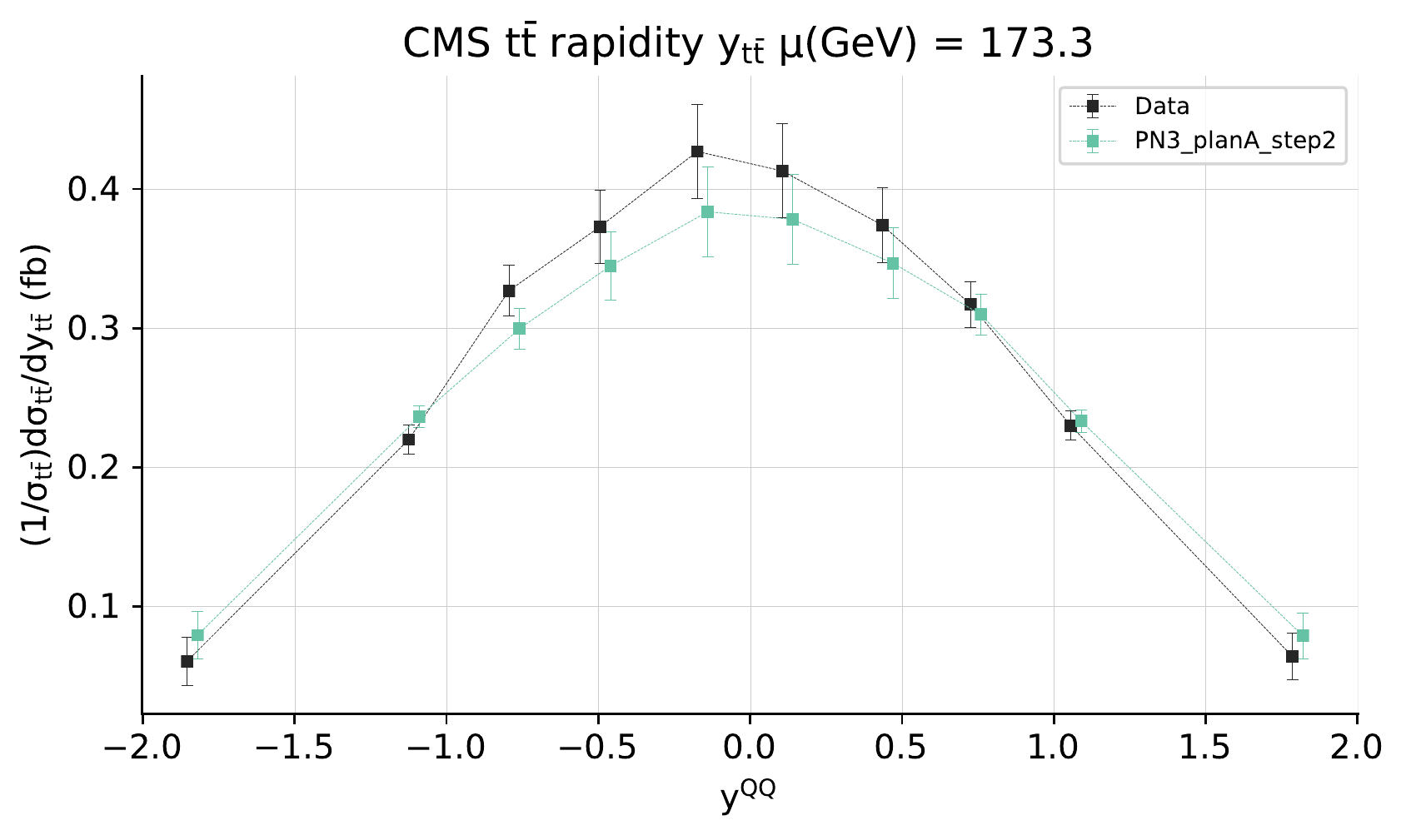}
    \caption{Data for the proton structure function $f_2$ measured at
      HERA (left) and top-pair production measured at the LHC (right)
    compared to a prediction based on PDFs determined from a fit to
    pre-HERA data\cite{NNPDF4}.}
    \label{fig:qa3}
\end{figure}

We have thus produced a PDF determination using {\tt n3fit} methodology, but
only including pre-HERA data, and now performing a dedicated
hyperparameter optimization based on this restricted dataset. The
best-fit gluon determined in this way is compared to the current
best-fit gluon in Figure~\ref{fig:futureg}. Some subsequent data which
are sensitive to the gluon, specifically the proton structure function
$F_2$, which is sensitive to the gluon at small $x$, and top-pair
production at the LHC, which is sensitive to the gluon at medium-high
$x$, are compared to predictions obtained using this PDF set in
Figure~\ref{fig:qa3}.

It is clear that the test is successful. In the region $x\lesssim 0.15$,
where the gluon is currently known accurately thanks to HERA data, but
it is   extrapolated when only using pre-HERA data, the uncertainty
grows very large, yet the two fits are compatible within these large
uncertainties, and the new data are within
the uncertainty of the extrapolated
prediction. This is a highly nontrivial test of the generalizing power
of the hyperoptimized {\tt n3fit} methodology. Note also that this
provides us with a test of the stability of the hyperoptimized
methodology, in that it means that a methodology hyperoptimized to the
much larger current dataset leads to reliable results even when used
on the much more restrictive past dataset.

The optimization of  the generalization power of our methodology is at
the frontier of our current understanding and remains a challenging
open problem.

\subsection{Outlook}

The {\tt n3fit} methodology will be used in the construction of future
PDF releases, starting with the forthcoming NNPDF4.0 PDF set. The
greater efficiency of this methodology will be instrumental in dealing
with an ever increasing data set, while its greater accuracy will be
instrumental in reaching the percent-level uncertainty goal which is
likely required for discovery at the
HL-LHC~\cite{Azzi:2019yne}. Avenues of research for future
methodological developments which are currently under consideration
include the possibility of an integrated reinforcement learning
framework for the development of an optimal PDF methodology, the
exploration of machine learning tools alternative to neural networks,
such as Gaussian processes,  the exploration of inference tools,
such as transfer learning, for the modeling of theoretical
uncertainties, and a deeper understanding of the generalizing power of
the methodology outside the data region.

\section*{Acknowledgments}
We thank all the past and current members of the NNPDF collaboration
and the members of the N3PDF team, whose work is behind most of the
results reported here, for collaboration and for countless stimulating
discussions, and in particular Rosalyn Pearson for a reading of the
first draft.
We acknowledge financial
support from  the European
Research Council under the European Union's Horizon
2020 research and innovation Programme (grant agreement n.~740006).

\clearpage

\bibliographystyle{tepml}
\bibliography{parton}


\end{document}